\documentclass[fleqn,usenatbib]{mnras}

\usepackage[T1]{fontenc}

\DeclareRobustCommand{\VAN}[3]{#2}
\let\VANthebibliography\thebibliography
\def\thebibliography{\DeclareRobustCommand{\VAN}[3]{##3}\VANthebibliography}

\usepackage{graphicx}	
\usepackage{amsmath}	
\usepackage{amssymb}	

\usepackage{color}
\usepackage[dvipsnames]{xcolor}
\usepackage{ulem}
\DeclareGraphicsExtensions{.png,.pdf,.eps,.jpg,.tiff,.ps}

\newcommand{\vunit}{\mbox{\,km\,s$^{-1}$}}

\newcommand{\Msun}{\mbox{\,$M_\odot$}}

\newcommand{\Rsun}{\mbox{\,$R_\odot$}}

\newcommand{\mic}{\mbox{$\,\mu$m}} 
\newcommand{\rso}{\mbox{RS~Oph}}

\newcommand{\ltsimeq}{\raisebox{-0.6ex}{$\,\stackrel 
	{\raisebox{-.2ex}{$\textstyle <$}}{\sim}\,$}} 
\newcommand{\gtsimeq}{\raisebox{-0.6ex}{$\,\stackrel
	{\raisebox{-.2ex}{$\textstyle >$}}{\sim}\,$}}

\usepackage{graphicx}	
\usepackage{amsmath}	
\usepackage{amssymb}	
\usepackage{pdflscape}

\definecolor{cyan(process)}{rgb}{0.0, 0.72, 0.92}

\definecolor{amethyst}{rgb}{0.6, 0.4, 0.8}
\definecolor{azure}{rgb}{0.0, 0.5, 1.0}
\definecolor{byzantine}{rgb}{0.74, 0.2, 0.64}

\newcommand{\fion}[2]{\mbox{[\ion{#1}{#2}]}}

\title[Infrared observations of RS Oph 2021]
{Near-infrared spectroscopy of RS~Ophiuchi in 2021:
the calm, the storm, and the abatement}
\author[C. E. Woodward et al.]{C. E. Woodward,$^{1}$\thanks{E-mail: chickw024@gmail.com},
A. Evans,$^{2}$\thanks{E-mail: a.evans@keele.ac.uk},
D. P. K. Banerjee$^3$,
B. Kaminsky$^4$, 
S. Starrfield$^5$,
K. L. Page$^6$,\newauthor
R. M. Wagner$^{7,8}$\thanks{Deceased.}
\mbox{ } \\ \\
$^{1}$Minnesota Institute for Astrophysics, University of Minnesota,
116 Church Street SE, Minneapolis, MN 55455, USA \\
$^{2}$Astrophysics Research Centre, Lennard Jones Laboratory,
Keele University, Keele, Staffordshire,  ST5 5BG, UK\\
$^3$Physical Research Laboratory, Navrangpura,
Ahmedabad, Gujarat 380009, India \\
$^{4}$Main Astronomical Observatory, Academy of Sciences 
of the Ukraine, Golosiiv Woods, UA-03680 Kyiv-127, Ukraine\\
$^{5}$School of Earth and Space Exploration, Arizona
State University, Box 876004, Tempe, AZ 85287-6004, USA \\
$^{6}$School of Physics and Astronomy, University of Leicester, University Road, Leicester LE1 7RH, UK \\
${^7}$Department of Astronomy, The Ohio State
University, 140 W. 18th Avenue, Columbus, OH 43210, USA \\
$^{8}$Large Binocular Telescope Observatory, 933 
North Cherry Avenue, Tucson, AZ 85721, USA \\
} 

\date{Accepted XXX. Received YYY; in original form ZZZ}

\pubyear{2026}

\begin{document}
\label{firstpage}
\pagerange{\pageref{firstpage}--\pageref{lastpage}}
\maketitle

\begin{abstract}
We present near-infrared (NIR) observations of the 2021
eruption of the recurrent nova RS~Ophiuchi.
The dataset provides both pre- and post-eruption perspectives
on the eruption, as well as NIR spectra at high cadence.
The spectrum obtained in 2020 June
(14.3~years after the 2006 eruption, and 428.1~days before 
the 2021 eruption), is that of the red giant secondary,
on which are superimposed several emission lines 
which most likely arise in the red giant wind.
Spectra obtained during the eruption consist of emission
(including coronal) lines, superimposed on a 
bremsstrahlung continuum at 8900~K. The temperature
of the coronal gas is estimated to be $10^{6.0}$~K
on day 11.7, and $10^{5.9}$~K on day 31.7.
The high cadence observations, obtained on day~31.7 of the 
eruption, {provide no conclusive evidence for
rapid ($\ltsimeq1$~minute) variations in the \ion{He}{i} 
1.0833\mic\ line.} Data obtained about one year after the
eruption show that there {may have been changes in the 
spectral type of the secondary after the 2021 eruption.}
\end{abstract}

\begin{keywords}
circumstellar matter  --
stars: individual: RS~Ophiuchi --
novae, cataclysmic variables --
infrared: stars
\end{keywords}



\section{Introduction}

RS~Ophiuchi is the best studied of the 
{symbiotic recurrent novae (SyRNe)}. It is a semi-detached
binary  consisting of a M0/2\,III red giant (RG) secondary with
mass $M_{\rm RG}=0.68-0.80$\Msun, and a white dwarf (WD) primary
with mass $M_{\rm WD}=1.2-1.4$\Msun; its orbital period is
453.6~days \citep[see][for details]{brandi09}. The orbital
inclination is estimated to be $i=49^\circ-52^\circ$, while 
$a\sin{i}=0.77$~AU, where $a$ is the semi-major axis 
\citep{brandi09}. 
Using the formulae in \cite{eggleton83}, the radius 
$R_{\rm RL}$ of the RG's Roche lobe is $7.0\times10^{12}$~cm
for these orbital parameters.

The RG has a wind, with estimated mass-loss rate
$\dot{M}\simeq10^{-7}$\Msun\,yr$^{-1}$ and terminal velocity
$|V_{\rm w}|=18$\vunit; these values are based on 
observations of narrow H$\alpha,\beta$ emission
lines over an orbital period \citep*{somero17}. 
The wind velocity determined by \citeauthor{somero17}
is consistent with that for field M2 giants \citep*{wood16},
for which the stellar radius $R_*=60.5$\Rsun\ 
\citep[$4.2\times10^{12}$~cm;][]{vanbelle99}.
{This is comparable with the equivalent radius of the RG's
Roche lobe; however, whether or not the RG in symbiotic systems
fills the Roche lobe} remains a matter of debate 
\citep[see, e.g.][]{boffin25,merc25}.

Following the 2006 eruption,
emission by silicate dust was reported by \cite{evans07c}
and \cite{woodward08}. The mass-loss rate inferred by the
presence of silicates is $1.0-1.7\times10^{-7}$\Msun\,yr$^{-1}$ 
\citep{rushton22}, consistent with the optical spectroscopy
reported by \citeauthor{somero17} The presence of a strong
18\mic\ silicate feature indicates that it is not
newly-formed: it has been in the \rso\ circumstellar
environment for a considerable time and has survived
many SyRN eruptions.

\rso\ is known to have undergone eight previous eruptions,
in 1898, 1907(?), 1933, 1945(?), 1958, 1967, 1985, and 
2006, those marked ``(?)'' being possible eruptions 
\citep{anupama08}. The 1985 \citep{bode87} and 2006 
\citep{evans08} eruptions were particularly well 
studied, and observed over a wide wavelength range, from
X-rays to radio.

The eruptions are the consequence of a thermonuclear
runaway (TNR) on the surface of the WD, following mass
transfer through the inner Lagrangian point on to the 
surface of the WD. There has been some discussion as
to whether accretion onto the WD occurs via Roche 
lobe overflow \citep[][]{azzollini23} or
via the RG wind \citep*[e.g.,][]{walder08,bollimpalli18}.
When conditions at the base of the accreted layer are
suitable, the TNR ensues, leading to the eruption and
the  explosive ejection of $\sim10^{-7}-10^{-6}$\Msun\
of material, at several 1000\vunit.

The ejected material encounters the RG wind, shocks it,
and a reverse shock is driven into the ejecta. These 
events heat the ejected gas and RG wind to temperatures
approaching $10^6$~K, giving rise to coronal line emission
in the ultra-violet, optical, and infrared, and to X-ray 
emission \citep{azzollini23}. The shocks also result in
particle acceleration, and the production of non-thermal
radio emission \citep*{chomiuk21}. In many
respects, observing the eruption of a SyRN like \rso\ is 
like observing the evolution of a supernova remnant, 
but in fast-forward. {The fact that the WD mass
\citep[estimated to lie in the range 1.2--1.4\Msun;][]{brandi09}
is close to the Chandrasekhar limit,
and is of CO type \citep{mikolajewska17}, 
makes \rso\ particularly interesting from a Type~Ia supernova
progenitor perspective, and hence all the more fascinating.}

The eruptions of 1985 \citep{evans88} and 2006
\citep*{das06,evans07a,evans07b,evans07c,banerjee09}
were intensively observed in the infrared.
We present here infrared spectroscopy of the 2021 eruption,
from roughly one year before, to roughly one year after, 
the eruption.

\section{The 2021 eruption}  \label{rsoph-2}

{The 2021 eruption was discovered independently by 
K.~Geary (August 8.93 UT) and A.~Amorim
\citep[August 8.91;][]{geary21}; see \cite{beck21} for a 
compilation of early
observations. \cite{munari21} estimated that the eruption
occurred on 2021 August 08.50 (MJD~59434,50).
We take this to define the origin of time, $t_0$.
The visual magnitude at discovery was 5.0.}

The eruption was observed over the entire electromagnetic
spectrum, including $\gamma$-rays 
\citep{cheung22, abe25, phan25}, X-rays 
\citep{page22, orio22, orio23, ness23, islam24}, and
radio \citep{munari22r, deruiter23, lico24, nayana24}.
Comprehensive optical spectroscopy is described by 
\cite{munari21,munari22o} and \cite{tomov23}. 
These observations showed outflows with velocities
up to 3500\vunit. 

Optical spectropolarimetry is given by \cite{nikolov23}, 
who concluded that dust was present in the 
\rso\ system as early as day~2 of the eruption, but was 
destroyed between days~2 and~9. However, (presumably 
newly-formed) dust appeared at $\gtsimeq80$~days into
the eruption. If this is the case, \rso\ becomes the 
second SyRN in which dust formation occurred during an 
eruption \citep[the first being V745~Sco;][]{banerjee23}.
\citeauthor{nikolov23} found that, at all times,
the dust in \rso\ was asymmetrically  distributed.

The X-ray evolution of the 2021 eruption is described
by \cite{page22}. The Super Soft Source (SSS) phase first 
appeared on day~20.6, and a soft component was
clearly present on day~26.3; it persisted until at least
day~63. X-ray observations of the 2006 eruption revealed
35~s quasi-periodic oscillation (QPO) during the SSS
phase \citep{beardmore08, nelson08, osborne11}.
These QPOs were also seen during the 2021 eruption, 
although the X-ray evolution was quite different in the 
two eruptions \citep{page22}. The QPOs were present
from days 36.9--61.6 during the 2021 eruption.

We assume reddening $E(B-V)=0.73$ \citep{snijders87},
and the extinction law of \cite*{cardelli89} with a 
total-to-selective extinction value of 3.1. 
Historically, the distance of \rso\ has been taken to 
be $D\simeq1.6$~kpc \citep{bode87,barry08}. More 
recently, however, \cite{schaefer22} has given a
somewhat larger distance ($2710^{+198}_{-135}$~pc) 
based on Gaia data. We assume 2.71~kpc here.

\section{Observations}

\subsection{IRTF\label{irtf}}

Spectra of RS Oph were obtained with SpeX \citep{rayner03} 
on the 3.2~m NASA Infrared Telescope Facility (IRTF). 
The SpexTool pipeline \citep*{cushing04} was used to
reduce these data, and to apply corrections for 
telluric absorption(s). The spectra were flux calibrated 
using an A0 standard star (HD 171149) observed at a 
comparable airmass ($\Delta_{\rm AM}\ltsimeq0.15$). The
IRTF flux calibrations are accurate to $\pm10$\%. 
The resolving power corresponds to a velocity of
250\vunit. The observational log is presented in Table~\ref{obs}. 

We use 
\begin{equation}
T_{\rm conj} = \mbox{MJD~} 45\,043.04 + 453.6E \:\:,
\end{equation}
where $T_{\rm conj}$ is the epoch of RG conjunction
\citep{brandi09}, to determine the orbital phase for
our observations (see Table~\ref{obs}). Note that all
but one of our observations were obtained close to
quadrature; the exception is the observation of 
2023 April~22, obtained when the RG was at inferior
conjunction. We therefore expect that, with the
exception of the observation of 2023 April, the
RG will be subject to some irradiation by the WD
during the quiescent observations,  

{SpeX SXD\_K short spectra at high cadence using a 
$0.5\farcs\times15\farcs$ slit also were obtained on
day~31.7, shortly after the commencement of the SSS phase.
All spectra were obtained in the ''A-beam'' 
(no ABBA nodding), and were reduced through the standard SpeX 
pipeline with the AOV star HD~171149 
as the telluric calibrator. The first seven time series spectra
had a cadence of $\sim14.4$~s (consisting of a 1.853~s exposure,
3~coadds, detector read-out, and file write to disk), while the 
40~spectra were obtained with a $\sim6.0$~s cadence (single 
1.853~s exposures plus overheads). The airmass for the first and
last  of the latter set differed by only 0.005, the entire set
being obtained at airmass 1.150. During the high cadence
observations, there were some clouds but it was clear in the
direction of \rso. The scientific motivation for obtaining
these data was to search for short period changes in line
intensity, or velocity modulated effects 
(see Section~\ref{scadence}).}

\begin{figure*}
\centering
\includegraphics[width=1.3\columnwidth]{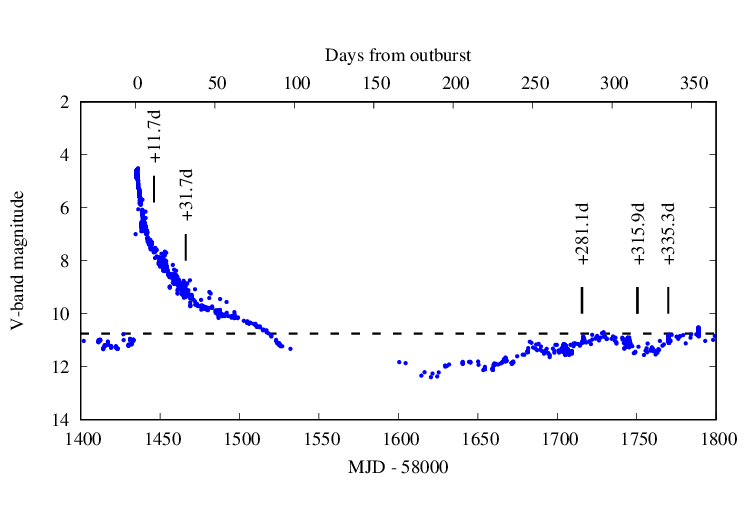}
\caption{AAVSO $V$ band light curve with the dates
(see Table~\ref{obs}) of the IRTF spectroscopy marked
(except for the days $-428.1$ pre- and  624.1 post- outburst).
In addition, spectra at high cadence were obtained on day~31.7
(see Section~\ref{scadence}).
The horizontal dotted line is the mean value of the pre-eruption
(MJD$<591200$) $V$ magnitude.
\label{LC1}}
\end{figure*}

The times of these observations in relation to the 
$V$-band light curve evolution\footnote{Obtained from the
AAVSO database https://www.aavso.org/} are shown in 
Fig.~\ref{LC1}.
Note the well-known dip in the light curve below the 
pre-eruption value around day~84, followed by eventual
recovery by day~300. This is likely due to the 
blowing away of the accretion disc (AD) by the eruption,
and its re-establishment by day~300 
\citep[see, e.g.,][for the resumption of accretion 
after $\sim200$~days following the 2006 
eruption]{bode06,worters07}.

\begin{table*}
\caption{IRTF SpeX Observational Summary. \label{obs}}
 \begin{tabular}{ccccccc}
UT        & MJD$^a$ & $t-t_0$    & $\Phi^c$ & SpeX & Resolving & SXD/LXD\\
Date/Time &   Days  & (days)$^b$ &          & Mode & power     &  ITOT$^d$ (sec)\\\hline
2020-06-10T09:37:05.663 & 59006.901 &--428.1 &0.783 & SXD/LXD & 1200 & 652/556\\
2021-08-20T04:52:18.456 & 59446.703 &11.7    &0.753 & SXD/LXD & 1200 & 178/574\\
2021-09-11T05:17:49.785 & 59466.721 &31.7    &0.797 & SXD     & 1200 & 311/\ldots \\
2022-05-16T13:24:50.957 & 59716.059 &281.1   &0.347 & SXD/LXD & 1200 & 593/566 \\
2022-06-20T09:41:46.586 & 59750.904 &315.9   &0.424 & SXD/LXD & 1200 & 593/723 \\
2022-07-10T08:15:43.383 & 59770.344 &335.3   &0.466 & SXD     & 1200 & 949/\ldots \\
2023-04-24T13:35:45.014 & 60058.566 &624.1   &0.103 & SXD/LXD &  750 & 297/834\\
  \hline\hline
\multicolumn{7}{l}{$^a$Julian Date $-2\,450\,000.$}\\
\multicolumn{7}{l}{$^b$Day $0\equiv t_0 = 59434.50$, 
  2021 Aug 08.50 ($\pm0.01$).} \\
 \multicolumn{7}{l}{$^c$Phase using ephemerides of \cite{brandi09}.}\\
  \multicolumn{7}{l}{$^d$Total integration time.}\\
  \end{tabular}
\end{table*}

\subsection{Swift}
\rso\ was observed with the UV/Optical Telescope (UVOT)
on the Neil Gehrels Swift Observatory
\citep{gehrels04}. The data closest to our IRTF observations
were obtained in the $uvm2$ (2246\AA) filter. The Swift 
photometry is given in Table~\ref{uvot};
the UVOT magnitudes were converted to fluxes using the 
calibration given by \cite{breeveld11}.
The fluxes in Table~\ref{uvot} are not dereddened.

\begin{table*}
\caption{Swift UVOT photometry.\label{uvot}}
 \begin{tabular}{cccccc} 
  MJD & $t-t_0$ & $uvm2$ & Flux & $uvw2$ &  Flux\\  
  Days & (days) & (2221\AA) & ($10^{-14}$ W m$^{-2}\,\mu$m$^{-1}$)   
  & (1191\AA) & ($10^{-14}$ W m$^{-2}\,\mu$m$^{-1}$) \\\hline
 59466.36 &  31.86 & $<10.42$ &  $>314$ & $<10.95$ &  $>222$ \\
  59722.03 & 287.53 & $15.04\pm0.03$ & $4.49\pm0.11$ & --- &  ---\\
   59750.44 & 315.94 & $14.54\pm0.03$ &$7.07\pm0.17$ & --- & ---\\
  59782.23 & 347.75 &   $14.74\pm0.03$ & $5.92\pm0.15$& ---& --- \\\hline\hline
 \end{tabular}
\end{table*}

\section{Results}

\subsection{Quiescence: pre-eruption\label{q-pre}}

\begin{figure}
 \includegraphics[width=8.5cm,keepaspectratio]{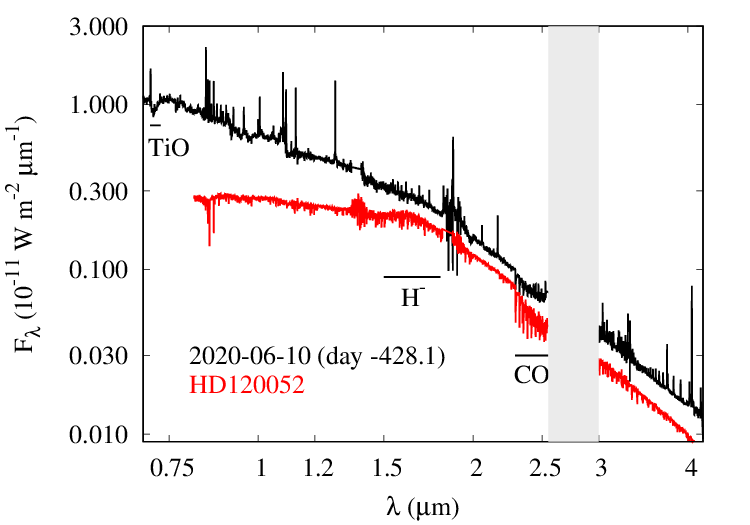}
 \caption{{Quiescent (i.e., pre-2021 eruption) 
 spectrum of \rso, obtained on 2020 June 10 (black). 
 Grey block indicates wavelength region
over which the atmospheric transmission is poor.
Spectrum of the M2III star HD120052, scaled to that of
\rso, is shown in red.
Locations of the $\gamma(0,0)$ TiO band (in \rso), the 
H$^-$ continuum, and of the first overtone CO bands
 are indicated. (cf. Fig.~\ref{post1}).}
  \label{quiescence}}
\end{figure}

A comprehensive optical (3900--7500\AA) spectroscopic
survey of \rso, with resolution up to 14000, was carried out
by  \cite{habtie25} over the period 2006--2021. Their
data suggest that the emission lines in their spectra arose in
the AD, and that a substantial change in the properties
of the AD occurred in the lead-up to the 2021 eruption.
They deduced an electron temperature for the AD of 
$T_e\simeq3.5\times10^4$~K, and an electron density $N_e$ in 
the range $\sim3.8\times10^9$~cm$^{-3}$
to $\sim3.4\times10^{10}$~cm$^{-3}$.
They also deduced a temperature and luminosity for the central
source, and a He abundance that is $\sim$~twice solar during the 
earlier period of their study (2008--2014), decreasing to the 
solar value at later times (2018--2020). 
While the time-span of their data has little overlap with
the data we present here, their study provides
useful pointers for interpreting the NIR data.

Our first observation was obtained on 2020 June 10, 
428.1~days before the 2021 eruption, and {5227.6~days}
($\sim14.3$~years) after the 2006 eruption (which occurred
on 2006 February 12.83 UT,  MJD 53778.83). We take the 2020
June observation to be close to ``quiescence'', and subtract
this spectrum from the two datasets obtained immediately after
the 2021 eruption. {This procedure is valid if we can be
sure that there were no variations between the pre- and 
post-eruption observations. Note, however, that the 2020
June observation was  obtained at the same orbital phase
as the two post-eruption spectra, which should minimise
the possibility of variability affecting the results.}

The quiescent spectrum of \rso\ is shown in 
Fig.~\ref{quiescence}, together with that of the M2III
field giant HD120052 from the
IRTF Spectral Library \citep*{cushing05,rayner09}.
It is apparent that the spectra of \rso\ and HD120052
are very similar longward of 1.8\mic, 
{although the first overtone CO bands seem somewhat
weaker in \rso.} However shortward of
1.8\mic, there seem to be two distinct differences. First,
the steep rise of the \rso\ spectrum to the blue;
this is likely due to the presence of the AD.
Second, the relative weakness in \rso\ of the H$^-$ continuum
around 1.6\mic. This is not surprising, given the low
binding energy ($\sim0.75$~eV) of H$^-$, and the harsh 
environment experienced by the RG photosphere. The quiescent
observation was carried out when \rso\ was 
in quadrature (see Table~\ref{obs}).

There are three possible sources for the few emission lines
present on day --428.1: the RG wind itself, the AD, and the
material that bears the silicate grains seen in the aftermath
of the 2006 eruption \citep{evans07c,woodward08}. 
In each case it is likely that the emitting material is
photo-ionised by the WD, and hard radiation from the inner AD.
A list of the more prominent lines is given in 
Table~\ref{all-data-tab}.
{We have estimated the line fluxes by fitting
one or more gaussians to the line profiles.
The fluxes listed in Table~\ref{all-data-tab}}
are for the strongest, central, component where 
the structure of the line is complex.

\subsubsection{The H recombination lines}
The flux in the H recombination lines enables us to
estimate the mass of emitting hydrogen if the lines
selected are optically thin. 
For a uniform medium, we have
\begin{equation}
 \frac{M_{\rm H}}{\Msun} = 8.98\times10^{17} \:\:
 D_{2.71}^2 \:\: \frac{f_n\lambda_n}{A_nn^2b_n} \:\: 
 \frac{T_e^{3/2}}{N_e}
 \:\: \exp\left[ -\frac{1.569\times10^5}{n^2T_e\mbox{\,(K)}} \right ]
  \label{Hmass} \:\:,
\end{equation}
where $N_e$ is in cm$^{-3}$ and $D_{2.71}\equiv1$ for \rso. 
In equation~(\ref{Hmass}),
$f_n$ and $\lambda_n$ are respectively the line flux
(in W~m$^{-2}$) and  wavelength (in \mic) of the line
originating in upper level $n$, and $A_n$ is the Einstein
spontaneous emission coefficient. The coefficients 
$b_n$ that describe departure from thermal equilibrium
are taken to be
\begin{eqnarray}
 b_n & = & 0.047 \:\: T_e^{0.25} \:\:  \exp [-{T_n}/{T_e}] \\
 && (5\le{n}\le10;~~~ 5000\le{T_e}\mbox{~(K)}\le3.2\times10^5) 
 \nonumber \\
 & \equiv & 1 ~~(T_e>3.2\times10^5\mbox{~(K)}) \:\:. \nonumber 
 \label{bn}
\end{eqnarray}
The form of $b_n$ for the lower $T_e$ range
has been obtained by fitting the $T_e$-dependence of the
various $b_n$ in \cite{baker38}. The $T_n$ values
differ for different values of $n$; for $n=8$ and 9, 
used here, $T_8=5145$~K, $T_9=4031$~K. 

A simple test of optically
thin-ness may be carried out on lines originating in
the same upper level $n$, for example Br$\epsilon$ 
(9--4) and Pf$\delta$ (9--5). In such cases the flux
ratios depend only on the wavelengths, $\lambda_n$,
and Einstein emission coefficients $A_n$ of the
respective lines, which we take from \cite{vanhoof18}'s
Atomic Line List 
(v3.00b5)\footnote{https://linelist.pa.uky.edu/newpage/}.

Table~\ref{all-data-tab} shows
that the flux ratio for these lines arising from 
$n=9$ is $\sim0.56$, close to the expected value
of $0.50$ for optically thin emission. We use this
line pair to estimate the mass of emitting H
in the quiescent spectrum.

As noted above, there are three possible sites for the H
emission lines: the RG wind, the AD, and the material that
bears the silicate dust \citep{evans07c,woodward08}
and we examine each case below:
\begin{enumerate}
\item The RG wind. If the H lines are produced in the RG wind,
we use an estimate of $T_e\sim8900$~K (see below). In
this case $M_{\rm H}/\Msun\simeq950/N_e$, with $N_e$ in
cm$^{-3}$. From the mass-loss rate and the time interval
$\Delta{t}$ between the 2006 eruption and our 2020 observation
($\simeq5227$~days), the wind mass is $\sim1.4\times10^{-6}$\Msun,
independent of the distance; this requires that 
$N_e\simeq6.6\times10^8$~cm$^{-3}$. The mass-loss rate from
the RG implies an electron density at the Roche lobe radius
of $\sim3.4\times10^9$~cm$^{-3}$. 
Given the approximations, this value of $N_e$ seems
consistent with the supposition that the H recombination
lines arise in the RG wind.
\item The AD. The mass of the AD may be estimated from 
$7\times10^{-10}[R_{\rm disc}/3\times10^{10}\mbox{cm}]^3$\Msun,
where $R_{\rm disc}$ is the outer disc radius \citep{wynn08}.
For a ``cold'' disc \citep[see][for details]{wynn08},
$R_{\rm disc}\sim10^{12}$~cm, giving a disc mass
of $2.6\times10^{-5}$\Msun; for a ``hot'' disc, 
$R_{\rm disc}\sim10^{11}$~cm and the disc mass is 
$2.6\times10^{-8}$\Msun. 
Using the $N_e$ and $T_e$ values from \cite{habtie25},
which apply to the AD, we find $M_{\rm H}$
in the range $1.2\times10^{-7}$\Msun\
to $1.2\times10^{-6}$\Msun.
As this range excludes the estimates of the AD mass,
an origin in the AD seems implausible, but
with the caveat that
we have implicitly assumed that the AD is optically thin.

However, an alternative estimate of $T_e$ comes from
the coronal lines \fion{Si}{vi} $\lambda=1.965\mic$
and \fion{Si}{ix} $\lambda=3.9357$\mic, both of which
are present in the 2020 spectrum (see 
Table~\ref{all-data-tab} 
{for data, and Section~\ref{clines} below for
methodology)}. The flux ratio in these lines
implies $T_e\simeq5.5\times10^5$~K, substantially
higher than the value found ($3.5\times10^4$~K) by 
\cite{habtie25}. The discrepancy is likely due to the
fact that their (optical) study did not include any
coronal lines, the presence of which inevitably inflates
the temperature estimate. This value of $T_e$ gives
$M_{\rm H}/\Msun \simeq1.5\times10^5/N_e$, with $N_e$
in cm$^{-3}$. The critical electron density for both
these Si lines, above which the upper level is 
collisionally de-excited, is $\sim10^9$~cm$^{-3}$ at 
$5.5\times10^5$~K. As this implies that 
$N_e\ltsimeq10^9$~cm$^{-3}$, this leads
to $M_{\rm H}\gtsimeq1.5\times10^{-4}$\Msun, 
considerably higher than the estimates of the AD mass.
We conclude that the H recombination lines cannot
arise in the AD.
\item \cite{woodward08} estimated the mass of silicate dust
seen in the Spitzer IRS data to be a model-dependent
$\sim2\times10^{-9}$\Msun. For a reasonable dust-to-gas
ratio ($\sim0.01$), the total mass of this material would be 
$\sim2\times10^{-7}$\Msun. If we take the same parameters as for
the RG wind, then $M_{\rm H}/\Msun\simeq950/N_e$.
In addition to the silicate features, the Spitzer IRS
data show the presence of the \fion{Ne}{ii} 12.81\mic\
and \fion{O}{iv} 25.91\mic\ fine structure lines
\citep{rushton22}. The critical densities for these
lines at $10^4$~K are $7.03\times10^5$~cm$^{-3}$ 
\citep{gehrz08} and $9.94\times10^3$~cm$^{-3}$ 
\citep{helton12} respectively. As these are upper
limits on $N_e$, the implied $M_{\rm H}\gtsimeq1.4\times10^{-3}$,
ruling out this interpretation.
\end{enumerate}
The H mass as inferred from the H recombination lines
therefore point to an origin in the RG wind, ionised
by hard radiation from the WD and inner AD.

There is evidence for neither a Paschen nor a Brackett 
discontinuity in the spectrum. The contribution of any 
free-free and free-bound emission for the case of a
pure hydrogen plasma may be set by the H line fluxes using
the {\sc nebcont} task in the STARLINK
\citep{currie14,bell24} package. Using the Pf$\delta$ line 
flux to set the level of the free-free and free-bound
emission leads to a continuum that is $\gtsimeq10$ times lower 
than that observed at the shortest wavelengths. 

\subsubsection{Metal lines}

Also present in the quiescent spectrum are \ion{O}{i}
fluorescent lines and the calcium triplet.

The \ion{O}{i} lines at 0.8446\mic\ and 
1.1288\mic\ are produced by the \cite{bowen47} fluorescence
mechanism. Each transition giving rise to a 1.1288\mic\ 
photon leads to the emission of an 0.8446\mic\ photon
\citep{rudy03}. This is essentially borne out by the
respective photon fluxes in Table~\ref{all-data-tab},
$\sim4.0\times10^4$~photons~m$^{-2}$~s$^{-1}$ (1.1288\mic)
and $\sim4.3\times10^4$~photons~m$^{-2}$~s$^{-1}$ 
(0.8446\mic). In their study of mass transfer in
T~CrB, \cite*{planquart25} concluded that these
\ion{O}{i} lines arise in the ``bright spot'', 
where the inter-star stream impacts the outer edge of
the AD. Given the broad similarity between the \rso\ and
T~CrB systems, this 
seems a likely interpretation of the \ion{O}{i} lines 
in the quiescent spectrum of \rso.

Also present in emission is the \ion{Ca}{ii} triplet
at 0.8500\mic,  0.8544\mic, 0.8664\mic.
{Allowing for the contribution of the Pa 11-3, 12-3, 14-3
lines to the Ca triplet (note that the flux values in 
Table~\ref{all-data-tab} include contributions from both
\ion{Ca}{ii} and Pa lines), the line \ion{Ca}{ii} 
fluxes are in the ratio 1:0.8:1.0}, close
to the ratio 1:1:1 for the case that the lines are 
optically thick; in the optically thin case the flux ratios
would be 1:9:5 \citep{herbig80}. These lines 
{have been detected in a number of cataclysmic
binaries by \cite{persson88}, who concluded that they
most likely arise in the AD. Given that the \ion{Ca}{ii} triplet 
in the case of \rso\ appears to be optically thick,
they likely arise, like the \ion{O}{i} lines, in either
the bright spot or the AD.}

\subsection{The 2021 eruption}

\begin{figure*}
 \includegraphics[width=1.5\columnwidth]{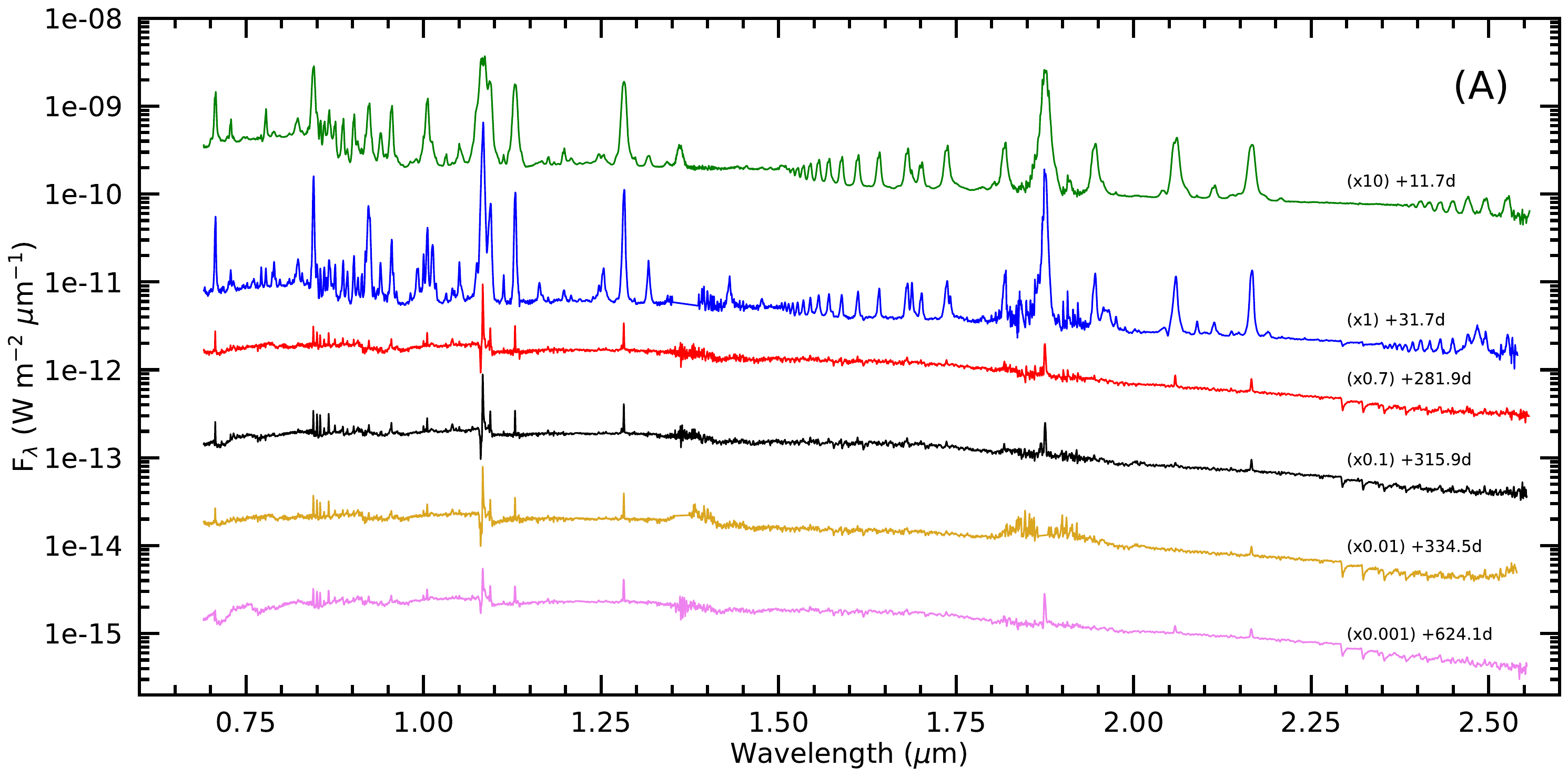}
 \includegraphics[width=1.5\columnwidth]{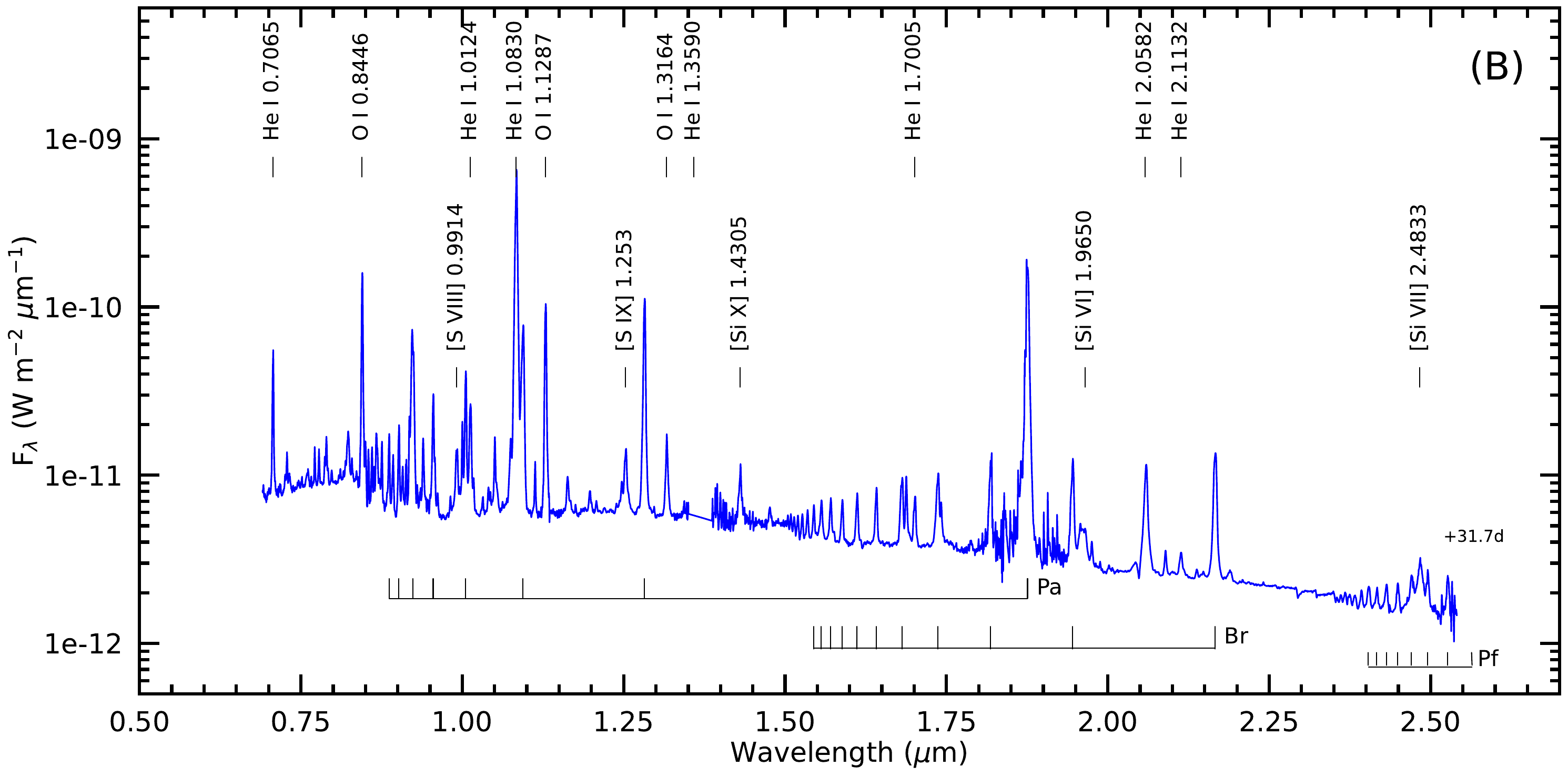}
 \includegraphics[width=1.5\columnwidth]{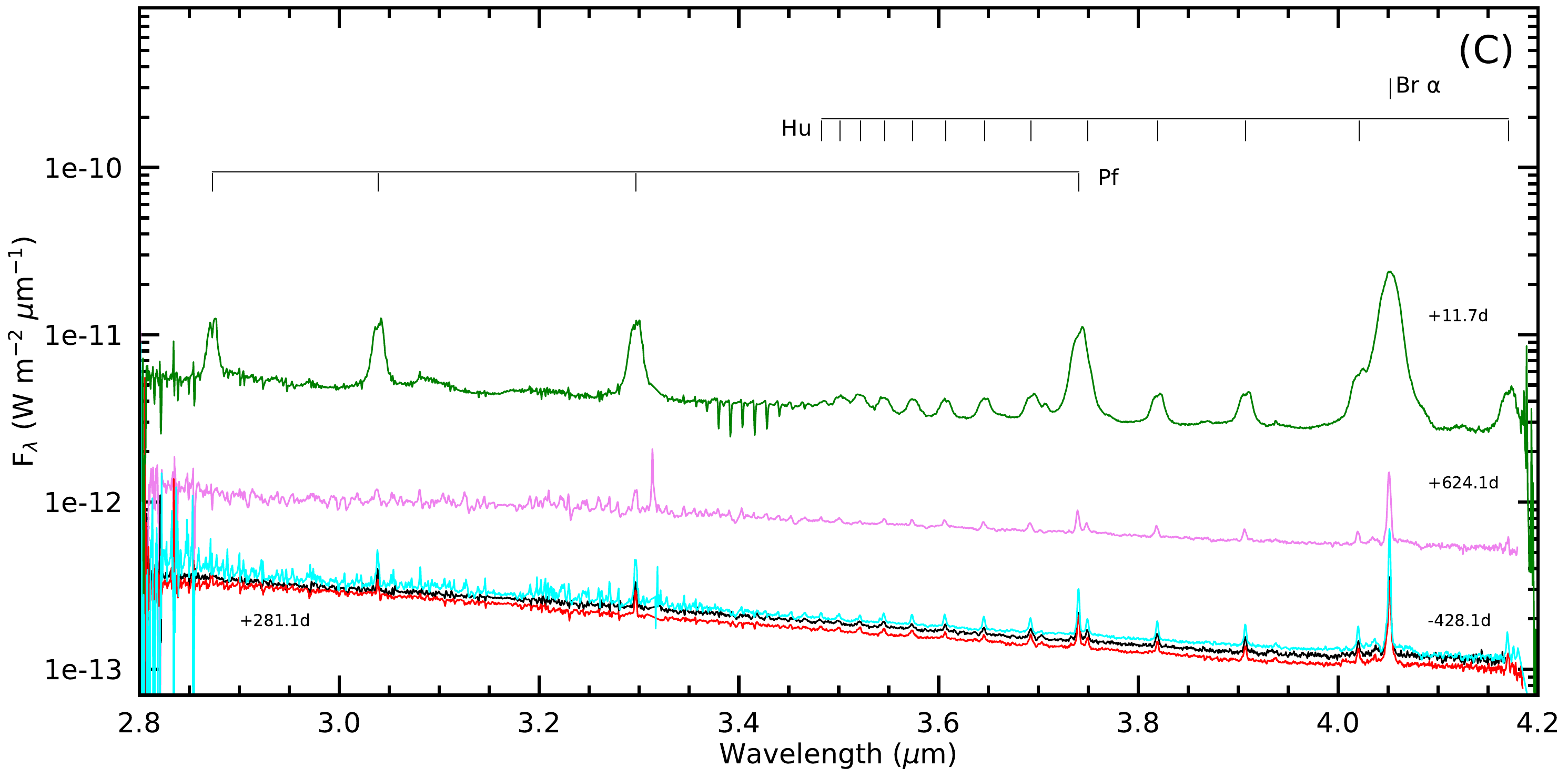}
   \caption{Observed near-infrared spectra of RS Oph at
   different epochs (Table~\ref{obs}). 
   (A):~Evolution of the SpeX SXD 0.7--2.5\mic\ range; data
   are offset by values stated in the figure for clarity. 
   (B)~SpeX SXD 0.7--2.5\mic\ spectra obtained on day~+31.7.
   Indicated in the panel are the wavelength of 
   various hydrogen emission line series (Brackett, 
   Paschen, Pfund). Representative He and O 
   lines are also identified, together with
   various infrared coronal lines. 
   (C):~Evolution of SpeX LXD mode (thermal wavelengths) 
   2.7--4.3\mic\ range. Days past outburst are indicated,
   together with the wavelengths of emission lines in the
   hydrogen Brackett, Humphries, and Pfund series.
   The absorption features around 3.4\mic\ on day +11.7
   are telluric.
    \label{alldata}}
\end{figure*}

The post-eruption spectra are shown in Fig.~\ref{alldata}.
The top and bottom panels cover the wavelength ranges
0.7--2.5\mic\ and 2.7--4.3\mic, corresponding to the 
SpeX Short-Cross Dispersed (SXD) and Long-Cross Dispersed
(LXD) wavelength ranges, respectively. The central
panel identifies a number of the stronger lines on
day~31.7.

\begin{figure*}
 \includegraphics[width=14cm,keepaspectratio]{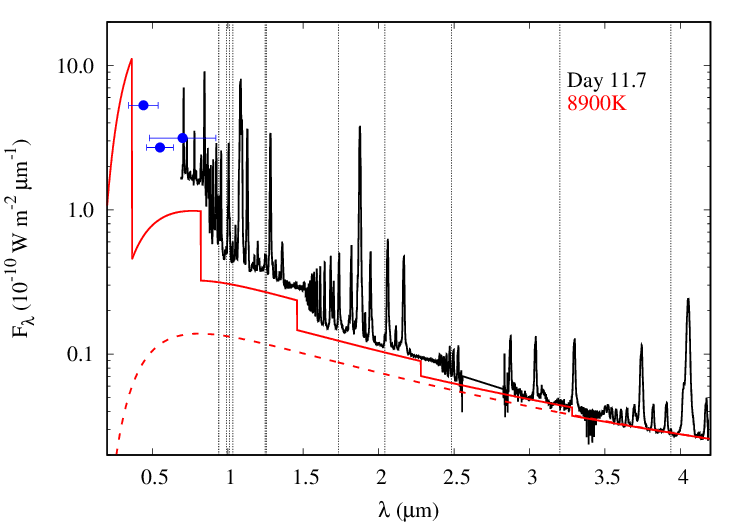}
  \includegraphics[width=14cm,keepaspectratio]{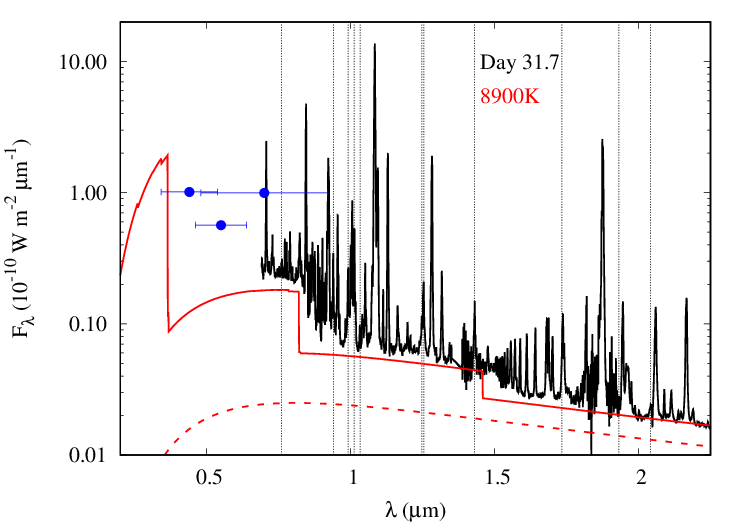}
  \caption{Top: IRTF spectrum for day 11.7;
  bottom: spectrum for day 31.7.
   Coronal lines are indicated by the vertical dotted lines
 (see Table~\ref{all-data-tab}).
  Data are shown in black. Full red curves are nebular continua
  for temperature 8900~K, for which the scale is set by H line fluxes. 
  Broken line in top frame is the free-free continuum
  for temperature 8900~K, for which the scale is set by fitting 
  the long ($\lambda>3.6$\mic) wavelength continuum.
  Broken line in bottom frame is the 8900~K free-free
  day~31.7 data as described in text.
  {Blue points are AAVSO $BV\!R$ data, for which the wavelength
  ``error bars'' are the FWHM of the $BV\!R$ filters.}
  See text for details.
 The plots are extended to 0.3\mic\ to show
 the expected extent of the Balmer discontinuity.\label{Day-117}}
\end{figure*}
The NIR spectra in the immediate aftermath of the 2021
eruption are shown in Fig.~\ref{Day-117}, together with
$BV\!R$ photometry from the AAVSO database.
{The AAVSO data were obtained within hours of our NIR spectra;
the photometric uncertainties as reported in the AAVSO data
are typically $\pm0.02$~mag (and are thus smaller than the plotted points in Fig.~\ref{Day-117}).} 

\begin{figure}
 \includegraphics[width=8.5cm,keepaspectratio]{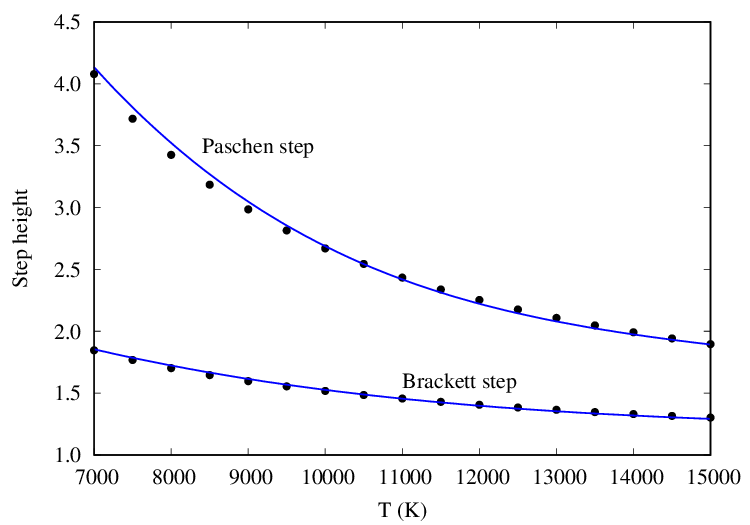}
 \caption{Dependence of the Paschen and Brackett 
 discontinuities on temperature for a pure hydrogen
 plasma. Black points: theoretical values  from 
 \protect\cite{baker38}; blue lines are polynomial fits.
 \label{disc}}
\end{figure}

The Paschen (0.821\mic) and Brackett (1.459\mic) discontinuities
are clearly visible in both day~11.7 and day~31.7 spectra.
Their magnitudes are suggestive of a relatively cool gas,
and \cite{evans07a} suggested that the material responsible
is the RG wind, ionised by the ultraviolet (UV) flash from 
the eruption \citep[see also][]{shore96}.
We explore this in more detail here by defining
the ``step height'' (SH) of the Paschen and Brackett 
discontinuities as
\begin{equation}
\frac{f_\lambda(\lambda\rightarrow0.821^-\mic)}{f_\lambda(\lambda\rightarrow0.821^+\mic)} 
\mbox{~~and~~}
\frac{f_\lambda(\lambda\rightarrow1.459^-\mic)}{f_\lambda(\lambda\rightarrow1.459^+\mic)}
\label{disc-rs}
\end{equation}
respectively. The dependence of these quantities on electron
temperature for a pure hydrogen plasma is
shown in Fig.~\ref{disc}. The magnitudes of
the discontinuities in \rso\ were determined by fitting 
the continuum blue-ward and red-ward of the discontinuity
with a quadratic function, after removing the emission
lines. we find $\mbox{SH(Pa)}\simeq2.56$,
$\mbox{SH(Br)}\simeq1.57$ (day~11.7),
and $\mbox{SH(Pa)}\simeq3.81$,
$\mbox{SH(Br)}\simeq1.68$ (day~31.7).
There is no evidence that the 
electron temperature changed between days 11.7 and 31.7,
and in the following we adopt a mean $T_e=8900\pm300$~K.

\subsubsection{Day 11.7}

The day~11.7 spectrum is shown in Fig.~\ref{Day-117}
(top panel), together with AAVSO $BV\!R$ data. 
At this time, the spectrum was dominated by H and He
emission lines, including Br 19-4 $\rightarrow$ 5-4, 
Hu 24-6 $\rightarrow$ 13-6, Pf 23-5 $\rightarrow$ 8-5 
series, and some of the Paschen series. Lines of carbon
(e.g., \ion{C}{i} 1.175\mic), nitrogen (e.g., \ion{N}{i}
1.247\mic), and oxygen (e.g., \ion{O}{i} 0.8446, 1.129,
and 1.3171\mic) were present, as is the \ion{Ca}{ii}
triplet at $\simeq0.86$\mic. The photon fluxes in the
Bowen-fluoresced \ion{O}{i} lines are again comparable,
$\sim1.7\times10^7$~m$^{-2}$~s$^{-1}$ (0.8446\mic)
and $\sim1.3\times10^7$m$^{-2}$~s$^{-1}$ (1.1288\mic).
The fluxes of the Ca triplet are in the ratio 1.0:0.8:2.3,
somewhat different from the ratios expected for the optically
thick case (see Section~\ref{q-pre}).

\begin{figure*}
\centering
\includegraphics[width=0.3\textwidth,keepaspectratio]{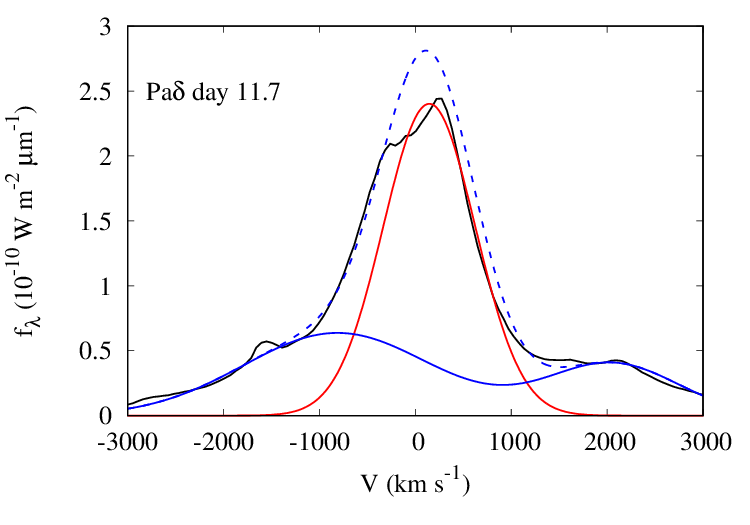}
      \includegraphics[width=0.3\textwidth,keepaspectratio]{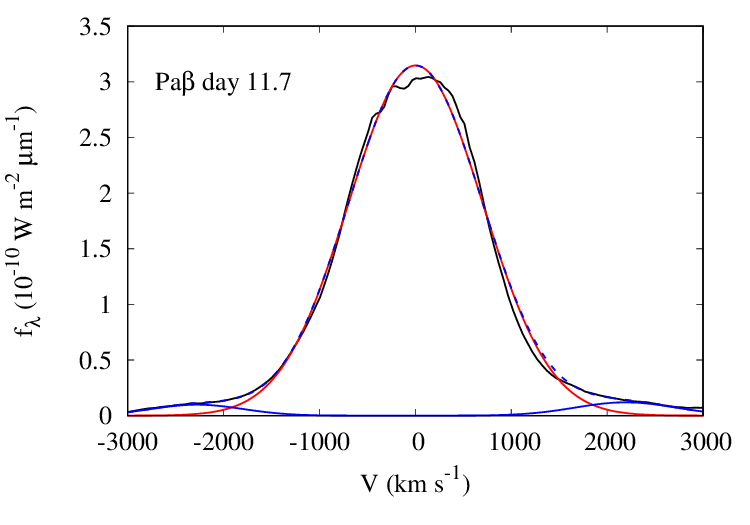}  
  \includegraphics[width=0.3\textwidth,keepaspectratio]{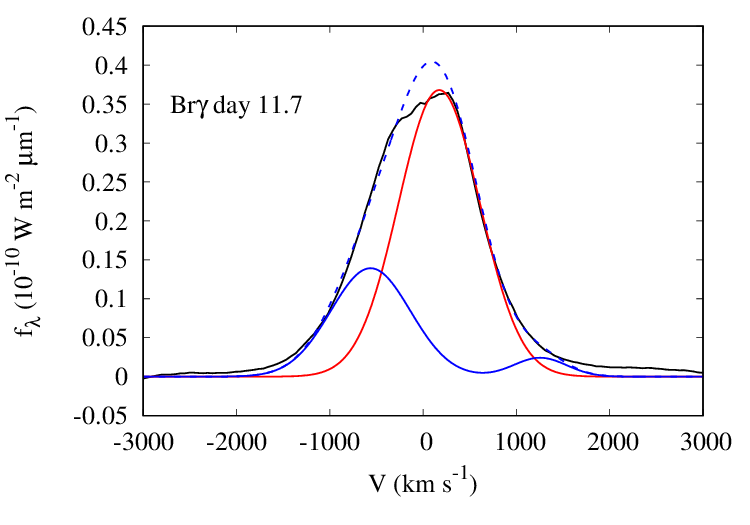}
     \includegraphics[width=0.3\textwidth,keepaspectratio]{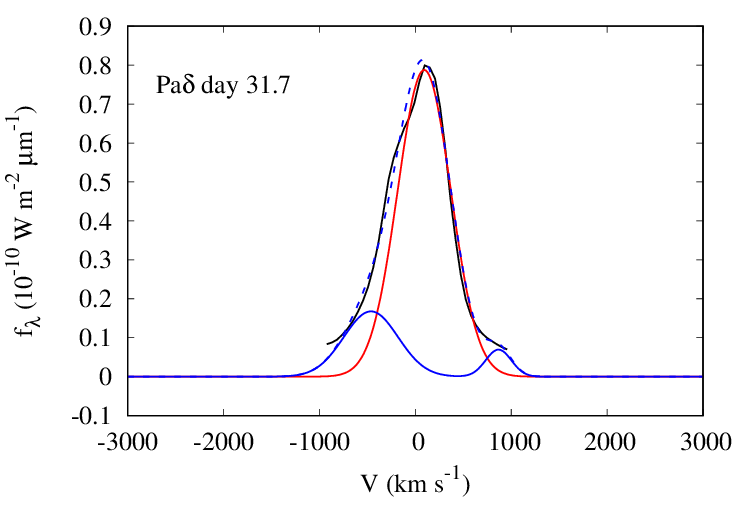}
     \includegraphics[width=0.3\textwidth,keepaspectratio]{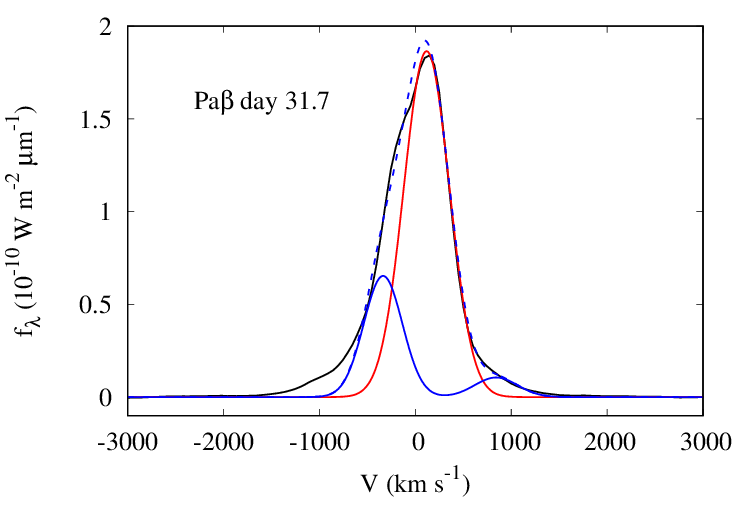}
\includegraphics[width=0.3\textwidth,keepaspectratio]{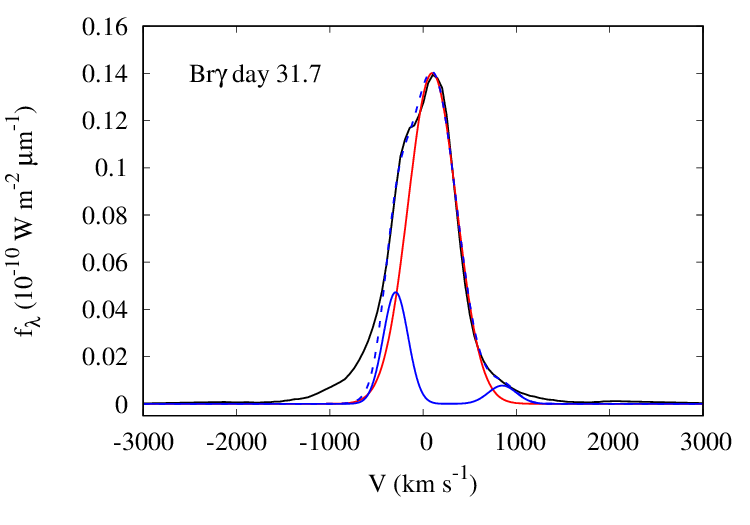}
     \caption{{Sample of H recombination lines profiles. 
     Black lines: continuum-subtracted data.
     Red lines: gaussian fit to line core.
     Blue lines: gaussian fits to line wings.
     Blue dashed lines: sum of core and wing fits.
     See text for details.}\label{hprofiles}}
\end{figure*}

The H lines consisted of a main peak, with wings to the blue
and red. The FWHM of the main peaks of the H lines, corrected
for spectral resolution, correspond to a mean expansion 
velocity of $\simeq 680$\vunit. The wings were displaced by
$\sim-1200$\vunit\ and $\sim+1900$\vunit, with the fluxes in
the blue wings being $\sim4$ larger than those in the red
{(see Fig~\ref{hprofiles}).}
A likely interpretation is that the main peaks arise in
a torus-like structure, while the wings arise in outflows 
perpendicular to the torus. Taking the orbital inclination
into account, and assuming that that the torus lies in the
orbital plane, the outflow velocities are $\simeq-650$\vunit\
and $\simeq+2200$\vunit.

\paragraph{Hydrogen emission:}
{For the purpose of the present paper, we
have computed the 8900~K nebular continuum (as implied
by the NIR discontinuities; see above)} for a plasma 
consisting of H and 10\% He by number \citep{habtie25}.
{A more thorough treatment of the continuum
and line emission will be given in a future paper.
In the {\sc nebcont} task, the}
scale of the nebular continuum is determined by the 
H$\beta$ flux, which is not available. We therefore use
the {flux in the central} component of the Br$\alpha$
flux for day~11.7, and the assumption of ``Case~B'', to
estimate the  H$\beta$ flux. This procedure 
{does of course assume} that the H recombination lines 
and the 8900~K nebular continuum arise in the same region,
and that Case~B applies, but we allow for this by scaling
the continuum by a factor $\ltsimeq5$. The result is
shown in Fig.~\ref{Day-117} (top panel), together with 
the dereddened AAVSO $BV\!R$ data. In addition to the Paschen
and Brackett discontinuities, the Pfund (2.279\mic) and
Humphries (3.282\mic) discontinuities are also evident.
{Our simple model satisfactorily reproduces the size
of the NIR discontinuities. However, we draw attention to
the expected large Balmer discontinuity, which was not
present in an optical spectrum obtained by \cite{munari21} within
four hours of our day~11.7 observation.}

{There may also
be some evidence for  an increasing discrepancy
between the data and the 8900~K nebular continuum as we go 
to shorter wavelengths. The AD, which was not re-established
until after day~225 \citep{munari22}, can surely make
no contribution at this early time.
A possible candidate is bremsstrahlung emission from
the hot coronal gas, but no combination of
8900~K gas and $10^{6.0}$~K gas 
(see Section~\ref{clines} below) can account for the
discrepancy. A similar shortfall was present in the 
day~11.81 spectrum of the 2006 eruption
\citep[see Figure~4 of][]{evans07a}. These factors
almost certainly highlight the complexity of the 
\rso\ environment compared with the simplicity of the 
model. A more spohisticated approach will be presented in
a future paper.}

{Notwithstanding the above caveats, we} use the
free-free emission to estimate the mass of
emitting H (note that the use of the H recombination 
lines for this purpose, as was done in Section~\ref{q-pre},
is complicated for spectra during eruption by the overlap
of several \ion{H}{i} and \ion{He}{ii} lines and possible
``contamination'' by lines from the still-undisturbed RG
wind). The free-free flux $f_{\rm ff}$
per unit wavelength interval is \citep[e.g.,][]{cox00}
{
\begin{eqnarray}
 f_{\rm ff} & = & 
 \frac{32\pi{e}^6}{3m^2c^3} \left (
\frac{2\pi{m}c^2}{3kT} \right)^{1/2}  \:\:
 \frac{N_e\mbox{(cm$^{-3}$)}}{4\pi{D}^2}
 \frac{1}{m_{\rm p}}
 \left ( \frac{M_{\rm H}}{\Msun} \right ) \nonumber \\
 &&\times  \:\: \overline{g}_{\rm ff} \lambda^{-2}
  \exp \left ( -\frac{hc}{\lambda{kT_e}} \right ) \:\:.
 \label{FF}
\end{eqnarray}
(in cgs units),} and $\overline{g}_{\rm ff}$ is the 
temperature-averaged
gaunt factor. The latter is virtually independent of wavelength
over the range 3.6--4.2\mic, where it has a value of 1.1
\citep{karzas61}; we take $T_e=8900$~K. Fitting the data
longward of the Humphries discontinuity (thus avoiding the
larger discontinuities at shorter wavelengths) gives 
\begin{equation}
 \frac{M_{\rm H}}{\Msun} \simeq \frac{207}{N_e\mbox{(cm$^{-3}$)}}\:
 \label{M-ff}
\end{equation}
for a distance of 2.71~kpc. This free-free continuum is shown
by the broken curve in Fig.~\ref{Day-117}.
By day~11.7, the shocked ejecta will have reached a distance 
\begin{equation}
 r_{\rm shock} = \frac{3}{2} V_0t_0 [t/t_0]^{2/3} \:\:,
 \label{rshock}
\end{equation}
\citep[appropriate for ``Phase~II'' of its evolution; see]
[for details]{bode85}
from the site of the explosion; here $V_0$ is the 
ejecta velocity at an early time $t_0$.
\cite{taguchi21} found $V_0=2600$\vunit\ one day after the 
eruption, so the shock radius is $r_{\rm shock} \simeq
3.4\times10^{11}[t/\mbox{days}]^{2/3}$~m 
$\simeq1.7\times10^{12}$~m on day~11.7. Note that the shock
would reach the outer edge of the wind, ``breakout'', at 
$t=138$~days in this case. The outer radius of the 
undisturbed wind at this time is
$\simeq8.1\times10^{12}$~m, so that the mass
of undisturbed (i.e., cool) wind is $\sim1.3\times10^{-6}$\Msun.
This implies $N_e\sim1.6\times10^8$~cm$^{-3}$,
but we should be mindful of the fact that the wind
is unlikely to be uniform.

\paragraph{Coronal lines:\label{clines}}
{There are several coronal lines in the day~11.7 spectrum. 
We use these, as in previous work 
\citep[e.g.,][]{woodward21,evans22},
to estimate the temperature $T_{\rm cor}$ of the coronal gas;
we defer the determination of abundances to a future paper.}
In general, the coronal lines (unlike the H recombination
lines) are unresolved, and so presumably arise in the same
region of the ejecta so that the $T_{\rm cor}$ may be
averaged. The ratio of the line fluxes for a pair $a$
and $b$ of ionisation states of the same element is
\begin{equation}
 \frac{f_a}{f_b} = \frac{n_a}{n_b} \:\: 
 \frac{\lambda_b}{\lambda_a}
 \:\: \frac{\Upsilon_a(T_{\rm cor})}{\Upsilon_b(T_{\rm cor})} 
 \:\: \frac{\omega_b}{\omega_a}
\end{equation}
\citep{greenhouse90}, where the $f$ values are the measured 
line fluxes, the ratio $n_a/n_b$ is a function of $T_{\rm cor}$,
and the $\omega$ values are the statistical weights of the lower
levels. 

{We use effective collision strengths from the IRON
project\footnote{http://cdsweb.u-strasbg.fr/tipbase/home.html}
\citep{hummer93, badnell06}. This dataset gives 
$\Upsilon$, the effective collision strengths, averaged 
over a Maxwell speed distribution, over a wide range
of temperatures; where the $\Upsilon$ values do not
go above temperature $10^5$~K, we use the value at $10^5$~K.
Ionisation fractions as a function of temperature are
determined using the data in \cite{arnaud85}.
\begin{table}
 \centering
 \caption{{Coronal lines used to estimate temperature of
 coronal gas. See text for details.}\label{tcor}}
 \begin{tabular}{clc}
  Day & Line pair & $\log{T_{\rm cor}}$ (K)\\ \hline
11.7 &  \fion{S}{xii}--\fion{S}{ix} & 6.13 \\
 & \fion{S}{viii}--\fion{S}{ix} & 5.76 \\
 & \fion{S}{ix}--\fion{S}{vi} & $>6.5$ \\
 & \fion{Si}{vii}--\fion{Si}{ix} & 5.92 \\
 & \fion{Al}{viii}--\fion{Al}{ix} & 6.15 \\
 & & \\
31.7 & \fion{Si}{vi}--\fion{Si}{x} & 5.89\\  
 & \fion{Si}{vii}--\fion{Si}{x} & 5.89 \\ 
 \hline\hline
 \end{tabular}
\end{table}
Using the ion pairs listed in Table~\ref{tcor}, we find a 
mean value  $\log{T_{\rm cor}}=
5.99\pm0.07$, or $T_{\rm cor} = 10^{6.0\pm0.1}$~K,
for day~11.7.}

\subsubsection{Day 31.7}\label{31.7}
The day~31.7 spectrum is shown in Fig.~\ref{Day-117}
(bottom panel), together with dereddened $BV\!R$ photometry
from the AAVSO database.
{The H lines again consisted of a main peak, with wings to the
blue and red (see Fig.~\ref{hprofiles}). However, there
was a significant change in that (a)~the velocity of the 
material from which the wings arise has decreased significantly,
from $-1200$\vunit\ to $-440$\vunit, and from $+1900$\vunit\
to $+850$\vunit, and (b)~the line widths had also decreased
substantially. This deceleration is a consequence of the
encounter of the ejecta with the RG wind, and was
first quantified for the case of
\rso\ by \cite{bode85}, who showed that the outer shock
radius at time $t$ is $r_{\rm shock}\propto{t}^{2/3}$
($V\propto{t}^{-1/3}$) during Phase~II of the 
development of the remnant.}

As for the day~11.7 data,
the 8900~K nebular continuum, scaled by the Br$\gamma$
flux, is included. 
It is evident in Fig.~\ref{Day-117} that, as on day~11.7,
the predicted nebular continuum falls short 
of matching the spectrum at the shortest wavelengths,
suggesting another source of continuum radiation.

The spectrum is again dominated by H recombination
and coronal lines. The Bowen \ion{O}{i} lines 
have photon fluxes $4.2\times10^6$~m$^{-2}$~s$^{-1}$
(0.8446\mic) and $3.2\times10^6$~m$^{-2}$~s$^{-1}$
(1.129\mic), while the flux ratios of the \ion{Ca}{ii}
triplet are 1.0:0.5:0.2.

\begin{figure*}
\includegraphics[width=1.6\columnwidth]{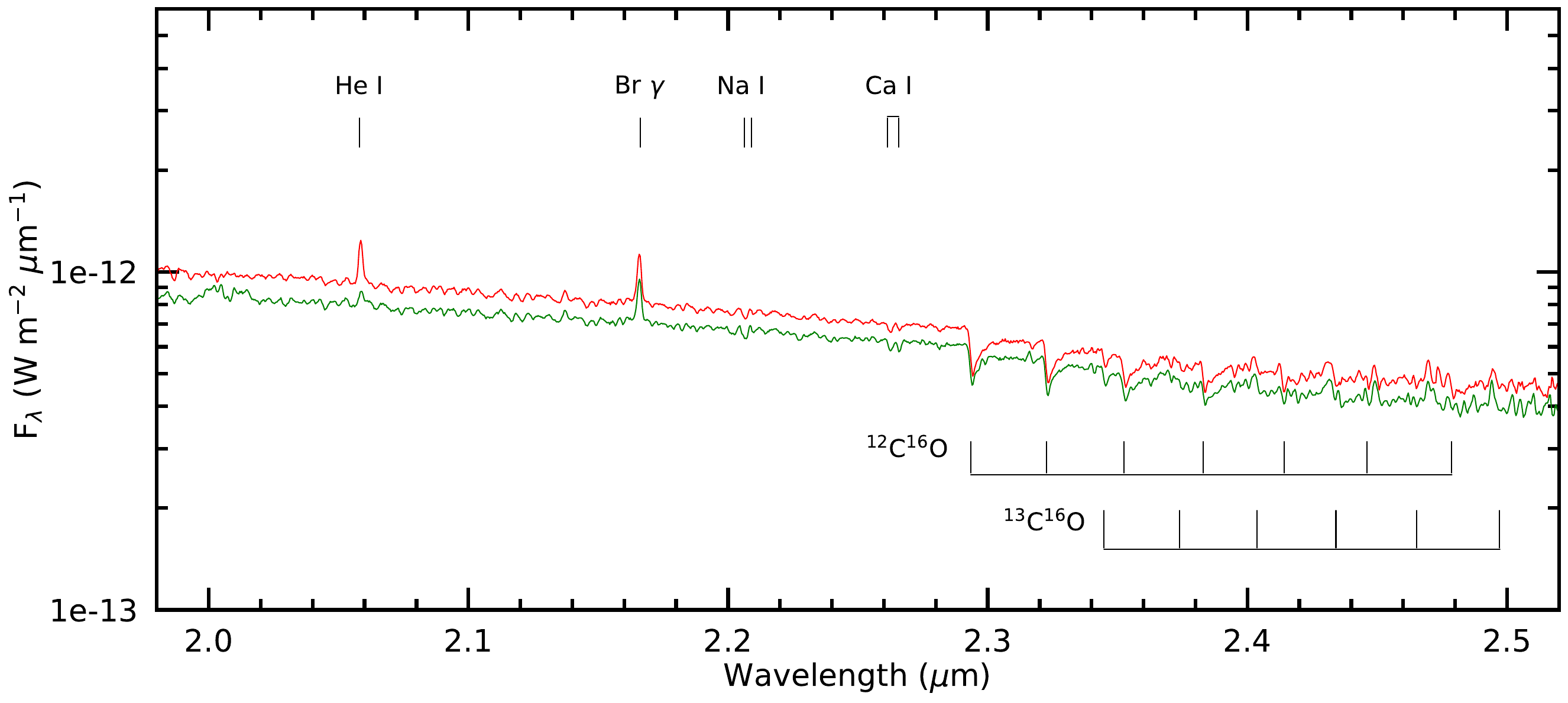}
\caption{The CO bandheads from the giant secondary as 
they emerge again to dominate the absorption features 
longward of 2.3\mic\ late in the post-outburst (2021) 
evolution of RS Oph.
Red: spectrum from 2022 May 16;
green: spectrum from 2022 June 20. \label{fig-cobandheads}} 
 \end{figure*}

The $^{12}$C$^{16}$O band heads from 2.29\mic{} onwards,
arising from the RG secondary, were clearly detected.
This is much earlier than in the 2006 outburst, when they
were not visible 55~days after the 2006 eruption
\citep{evans07a}, although the cadence of the NIR spectroscopy
was different for the 2006 and 2021 eruptions. 
The CO bandheads are quite distinct
$\gtsimeq 280$ days after the outburst as the system
returns to quiescence (see Fig.~\ref{fig-cobandheads}).
The CO bands are further discussed in Section~\ref{CObands}.

\paragraph{Coronal lines:}
IR coronal lines of \fion{S}{viii} 0.9914\mic\ (ionisation
potential of the lower ionisation stage 281~eV),
\fion{S}{ix} 1.253\mic\ (329~eV), \fion{Si}{x} 1.403\mic\ (351~eV), 
\fion{Si}{vii} 1.9650\mic\ (167~eV), and \fion{Si}{vii} 2.4833\mic\
(205~eV) appeared, just after the onset of the SSS on day 20.6
\citep{page22}. There are no data longward of 2.5\mic, and
we estimate the temperature of the coronal gas using the Si
lines \fion{Si}{x} 1.403\mic, 
\fion{Si}{vii} 1.9650\mic, and \fion{Si}{vii} 2.4833 \mic.
We find that $T_{\rm cor}=10^{5.9}$~K, not significantly different
from that on day~11.7. 

\paragraph{Hydrogen emission:}
The 8900~K continuum still provides a reasonable
description of the shape of the NIR continuum.
However in the absence of LXD data we can not, as we
did for the day~11.7 spectrum, fit a pure
free-free continuum to the data
using equation~(\ref{FF}).
However we scale the fit to day~11.7 by matching
the free-free flux at 2\mic\ to the nebular continuum.
The match is good if the day~11.7 free-free continuum 
is scaled by a factor 0.18; this continuum 
(again for 8900~K) is shown in the bottom panel of
Fig.~\ref{Day-117}. The implied mass of hydrogen is
$M_{\rm H}/\Msun\simeq37/N_e$.
By day~31.7, the outer edge of the undisturbed
wind is at $8.85\times10^{12}$~m, whereas the 
shock is at $3.38\times10^{12}$~m; the mass
of the undisturbed wind is $\sim9.6\times10^{-7}$\Msun.

The change in the estimated H mass from 
the free-free emission by the cool gas
likely rules out an origin in the silicate-bearing
material, of mass $\sim2\times10^{-7}$\Msun\
(see Section~\ref{q-pre}). It is likely that 
this material is unaffected by the violent events
occurring in the vicinity of the nearby binary.
By day~31.7, the FWHM of the H emission lines had 
narrowed (velocities of 610\vunit), and the red-wing
asymmetries and saddle-shaped line profile peaks had 
disappeared. 

\subsubsection{The high cadence observations\label{scadence}}

\begin{figure}
\centering
\includegraphics[width=0.515\textwidth,keepaspectratio]{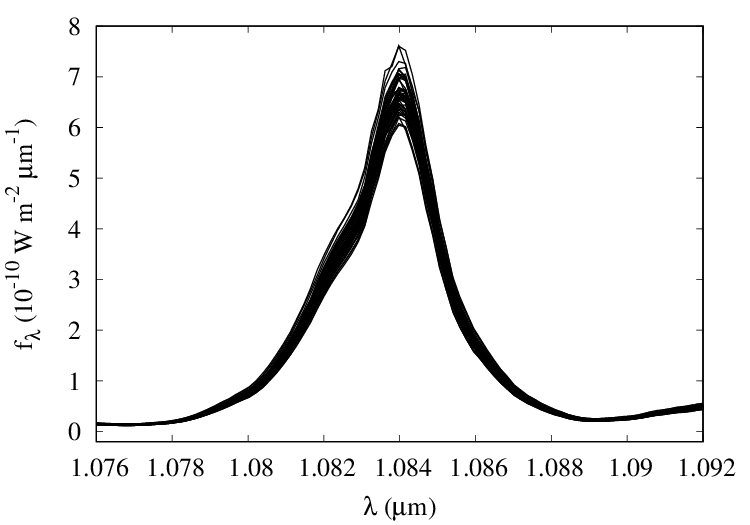}
\includegraphics[width=0.485\textwidth,keepaspectratio]{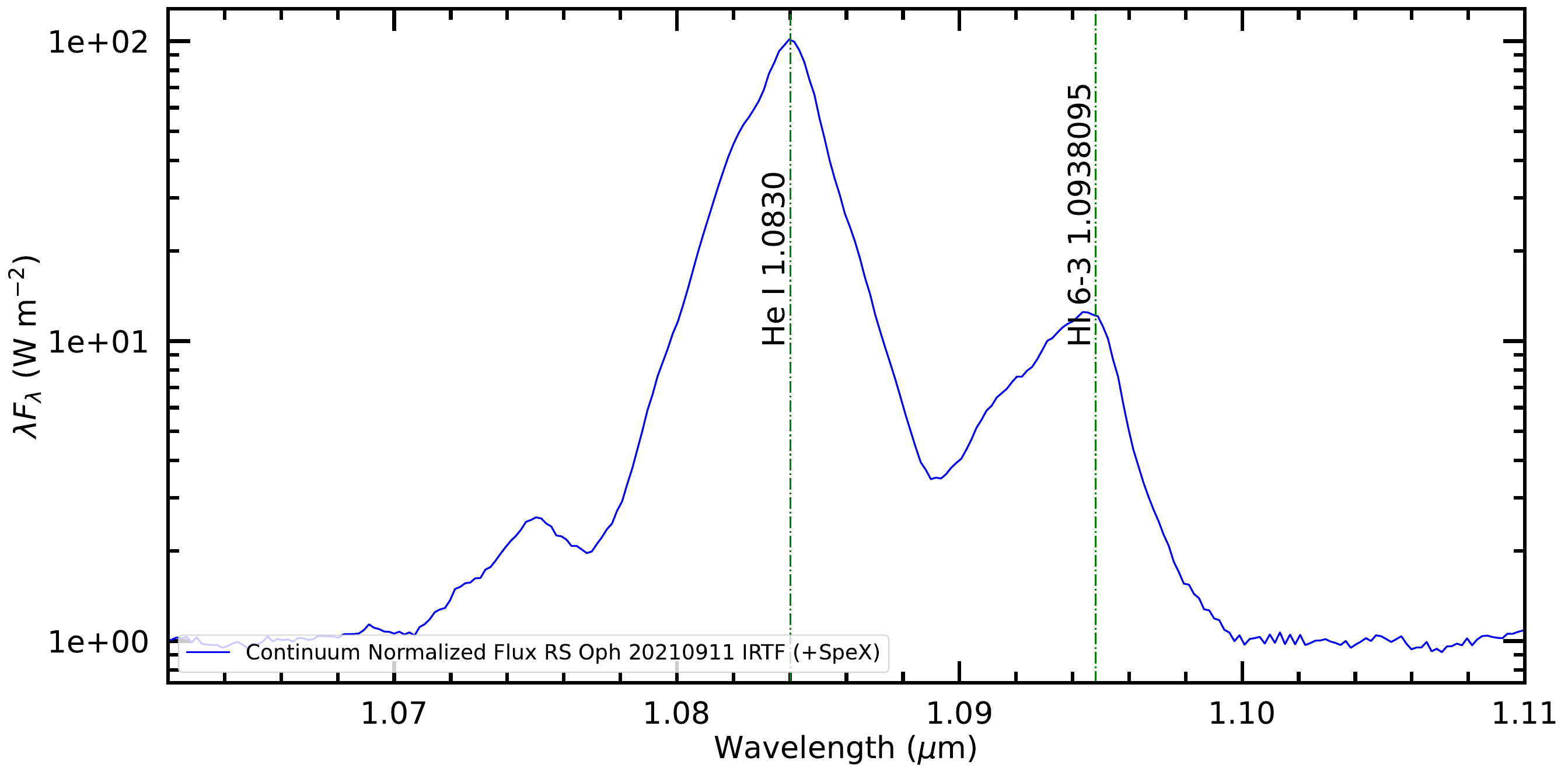}
\includegraphics[width=0.485\textwidth,keepaspectratio]{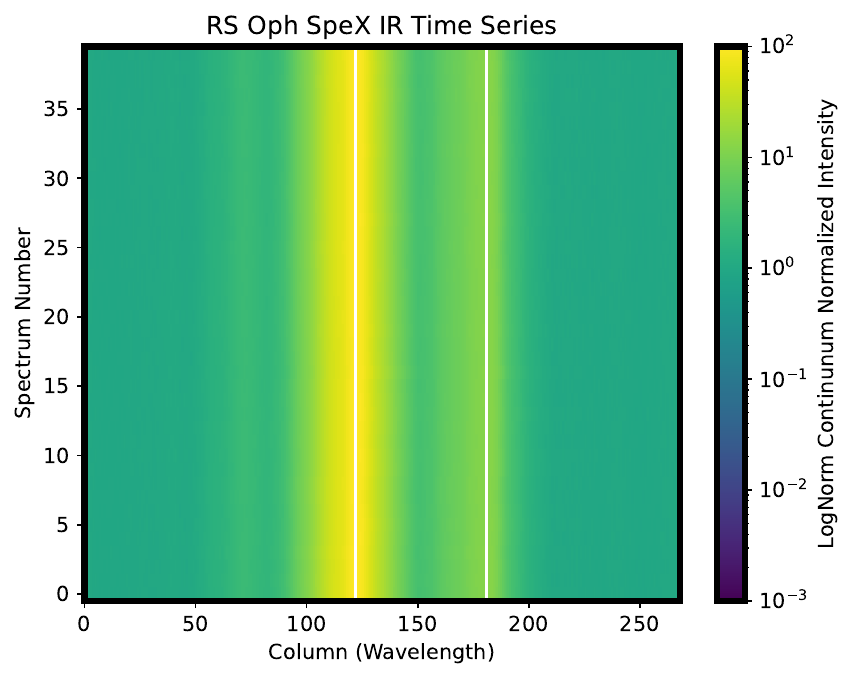}
\caption{{Top: Profile of the \ion{He}{i} triplet during high
cadence observations. {Fluxes are as observed,
and all high cadence data are shown.}
Middle:  Example of continuum normalised spectrum in the region of
interest near the \ion{He}{i} line. 
Bottom: Three-dimensional rendering of the time series spectra
obtained on a $\sim6$s cadence over a $\sim 240$s interval 
on day 31.5. The vertical white line at column 122, corresponds to the
wavelength of \ion{He}{i}, while the vertical
white line at column 181 is \ion{H}{i} 6--3 line at 1.0938\mic. 
The flux scaling uses a LogNormal rendering as indicated 
by the intensity bar to the right of the panel. Time scale 
starts at the bottom and increases upwards.}
\label{s-cadence}}
\end{figure}

The continuum around, and the profile and width of, the
\ion{He}{i} line show no {significant evidence of
variability during this time, as may be seen in the top 
panel of Fig.~\ref{s-cadence}.
Note that the variations in the peak flux in the line 
are within the expected 10\% uncertainty in the flux
calibration (see Section~\ref{irtf}).}
To further explore possible
variation in the \ion{He}{i} line on short time-scales, 
we extracted from the forty spectra obtained with the $\sim6$~s 
cadence a region of interest (ROI) from 1.0620\mic\ to 
1.1100\mic, computed the local continuum from a first-order
polynomial (python np.polyfit(wavelength, flux, 1), using
continua from 1.0620\mic\ to 1.0680\mic, and from
1.1020\mic\ to 1.1100\mic, creating continuum-normalised
spectra over the ROI (Fig.~\ref{s-cadence}, middle panel).
A three-dimensional representation of these data
is presented in Fig.~\ref{s-cadence}, bottom panel). There
is no clear evidence of marked variation in the \ion{He}{i}
1.0830\mic\ line over this period  of $\sim240$~s. In particular,
there is no evidence for any periodic variation
\citep[which might correspond, for example, to the 35~s
QPOs seen in the X-ray data;][]{page22}.

However, deeper synoptic observations, at higher cadence,
during the SSS phase of the next
eruption are required to confirm this
result. In particular, we plan similar
high cadence spectroscopy during the imminent
eruption of the SyRN T~CrB \citep[][{but see \cite{merc25+} 
for a cautionary note}]{schaefer23}.

\subsection{Quiescence: post-eruption}
\begin{figure*}
    \includegraphics[width=8.5cm]{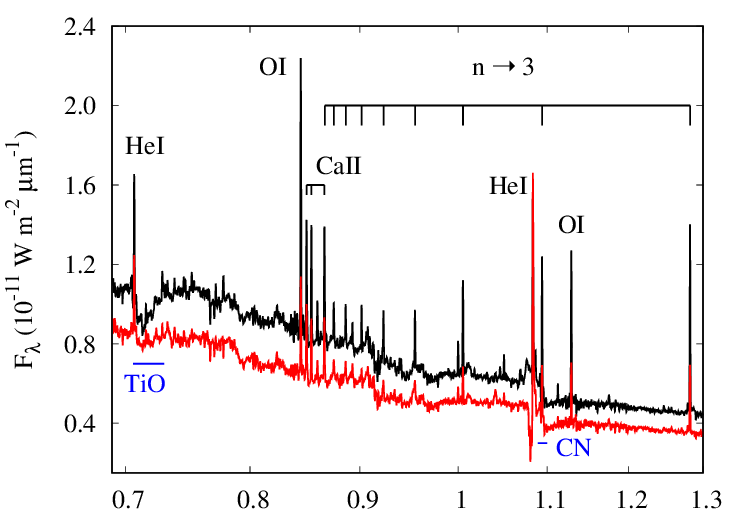}
   \includegraphics[width=8.5cm]{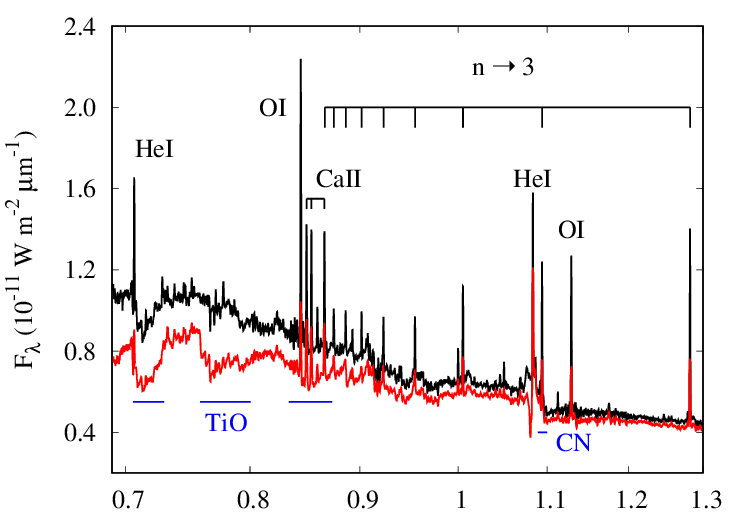}
  \caption{{Left: pre-eruption (day --428.1, black) and 
  post-eruption (day 335.3, red) spectra in the 
  0.69--1.3\mic\ range. 
  Right: pre-eruption (day --428.1, black) and post-eruption
  (day 624.1, red) spectrum. 
  In both panels, key emission lines and molecular features are identified, and
  the wavelength scales are logarithmic
  to stretch out the shortest wavelengths.}
  \label{post1}}
\end{figure*}

It is of interest to compare the pre-2021 eruption spectrum
with those obtained after the eruption had subsided.
Fig.~\ref{post1} shows the pre-eruption spectrum (day --428.1),
along with the spectra obtained on days 335.3 and 624.1 after
the 2021 outburst. 
In both of the later spectra, the post-eruption
continuum lies below the pre-eruption spectrum,
particularly in the blue on day~624.1; recall that the 
latter was obtained when the RG was at inferior conjunction,
so the visible hemisphere of the RG would not have been
affected by the WD. 

In the post-eruption spectra, the \ion{He}{i}
line at 1.0833\mic\ shows a prominent P~Cygni profile,
with terminal velocity $\sim-1500$\vunit.
The \ion{He}{i} line at 0.7067\mic\ is clearly present on
day 335.3. On day 624.1, it is weakly present; it too seems to have
a P~Cygni profile, with terminal velocity $\sim1000$\vunit,
but its presence close to the TiO band head makes drawing
any quantitative conclusion difficult.
The H recombination lines show no such structure.
The presence of P~Cygni profiles in the \ion{He}{i} lines,
but not \ion{H}{i}, suggests that, at this time, the He is 
located in a wind that still emanates from the WD, whereas
the H may be located in the ``permanent'' circumstellar
material associated with the \rso\ system.

The \ion{Ca}{ii} triplet, which was present in the 
pre-eruption spectrum (see Fig.~\ref{quiescence}), is 
also present in the quiescent post-eruption spectra,
as are the Bowen-fluoresced \ion{O}{i} lines.

\begin{figure}
 \includegraphics[width=8.5cm,keepaspectratio]{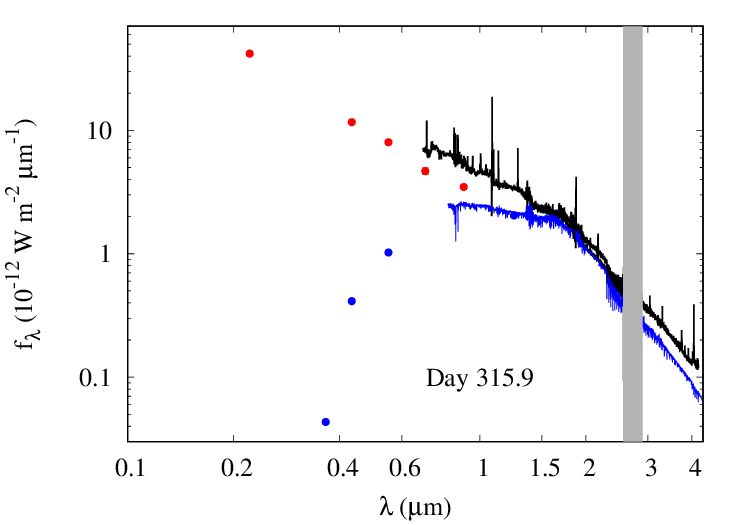}
 \caption{Day 315 spectrum of \rso\ (black curve), with Swift
 UVOT and $BV\!RI$ photometry (red points; error bars are smaller
 than the plotted points). \rso\ data dereddened by
 $E(B-V)=0.73$. Blue spectrum is that of HD120052,
 together with $U\!BV\!R$ photometry  (cf. Fig.~\ref{quiescence}).
 Grey block indicates wavelength region over which the 
 atmospheric transmission is poor.  \label{quiescence2}}
\end{figure}

There are simultaneous data (Swift UVOT, AAVSO $BV\!RI$, IRTF) for 
Day 315.9. As in Fig.~\ref{quiescence}, we compare the spectral
energy distribution (SED) of \rso\ with that of HD120052 
(Fig.~\ref{quiescence2}). There is an offset ($\sim60$\%) between
the photometry and the spectroscopy, likely due to the 
uncertainty in the flux calibration of the NIR spectrum.
However it is again evident that, as in the pre-eruption
quiescent spectrum (Fig.~\ref{quiescence}), there is a large 
discrepancy between the SED and that of the field giant in the 
blue-UV. We again attribute this to the presence of the AD,
which will have been re-established by day~315.9,
{and likely as early as day 225 \citep{munari22}}.

\subsection{The molecular bands}

\subsubsection{The TiO and CN bands}
\label{TiO}
{The three TiO bands at 0.705\mic\ ($\gamma(0,0)$),
0.759\mic\ ($\gamma(0,1)$), and 0.843\mic\ ($\epsilon$)
are present pre- and post-eruption,
but their prominence in each case differs considerably
(see Fig.~\ref{post1}).
The 0.705\mic\ band is formed in the outer layers
of the RG photosphere, the 0.759\mic\ band at somewhat deeper layers, and the 0.843\mic\ band at even deeper layers.}

{Only the 0.705\mic\ band is present
on day --428.1 (Fig.~\ref{quiescence}), and possibly on  
day~335.3. However it is evident that 
these three TiO bands are present on the last (day 624.1) spectrum.
In the spectra of S-type symbiotic stars 
described by \cite{murset99}, all three bands are present
in a M7.5 RG, but only the 0.705\mic\ band in
earlier (M0, M3.5) RGs. While irradiation of the RG by either 
the WD or the AD has negligible effect on the effective
temperature of the RG, the TiO bands in \rso\ may
suggest a change in the effective spectral type of
the RG secondary, before and after the eruption.
Indeed, in Fig.~\ref{post1}, it may be seen
that the overall slope of the SED for day 624.1
suggests a lower effective temperature.}

We do not consider that the changes
in the TiO bands are related to the 2021 outburst.
\cite{pavlenko16} presented six optical spectra, the
first of which was obtained 7~months after the 2006
outburst. All of these spectra, including the first,
have deep TiO bands, in particular the $\gamma'$ band
(band head at 0.620\mic). This suggests that the 2006
outburst did not affect the TiO bands, and we can
expect that this was also the case in 2021. There
is no significant variability in the TiO bands among
the spectra shown by \cite{pavlenko16}, but all those
spectra were obtained when the binary was 
approximately in quadrature.
In the present data, the day --428 spectrum
(Fig.~\ref{quiescence}) was also obtained close to 
quadrature, and the $\gamma(0,0)$ band is relatively
strong. On days 281, 315 and 335, when the binary was
also in quadrature, the spectra show weak TiO bands,
especially on day 335.3 (phase 0.466), when the bands
are weakest. On day~624.1, the RG was at inferior
conjunction, and there would have been no effect of the 
WD and AD; as expected, this is when 
the TiO bands are strongest. 

We considered the possibility that the TiO bands
might partially be formed in the shell, but we regard this
as unlikely. If this were the case, the presence of the
shell would also be appparent in the CO first overtone
bands.  However, our modeling of the high resolution
spectra of \rso\ in \cite{pavlenko10} showed that the
CO bands are reproduced quite well using a single model 
atmosphere, without additional layers.

{There are a number of possible sources of radiation
in the environment of the RG (e.g., the WD, the boundary
layer, the RG wind and still-present ejecta) which might
result in the ``dilution'' of the TiO bands. A detailed
consideration of this issue is complex, and beyond the
scope of this paper.}

Also detected are the CN bands $\Delta{v}=0$ R$_2$ 
(band head at 1.0932\mic), R$_1$ (1.0971\mic) and
Q$_1$ (1.1004\mic); these are indicated on the bottom
panels of Fig.~\ref{post1}. These bands are commonly
seen in the spectra of field RGs, and we therefore 
conclude that they originate in the RG component of \rso.

Weak 3\mic\ H$_2$O lines are also present on day~624, 
when the RG effective temperature was lowest; they  
may also be present on day~$-428$.

\subsubsection{The first overtone CO bands}
\label{CObands}
The $^{12}$C$^{16}$O band 
heads from 2.29\mic\ onwards, arising from the RG, 
were present in the day~31.7 spectrum
(see Section~\ref{31.7}). 
They were not present 55~days after the 2006 eruption
\citep{evans07a}, and were first detected on day 170
\citep{rushton10}. This behaviour during the 2021 
eruption is similar to that of the SyRN V3890~Sgr
(which also has a RG secondary), in which the CO bands were
clearly present  31.35~days after its 2019 eruption
\citep{evans22}. In \rso, the CO band heads are 
present on day~281.1 after the 2021 outburst,
as the system returns to quiescence (see
Figure~\ref{fig-cobandheads}).

\subsubsection{The first overtone SiO bands}

\begin{figure}
 \includegraphics[width=8cm,keepaspectratio]{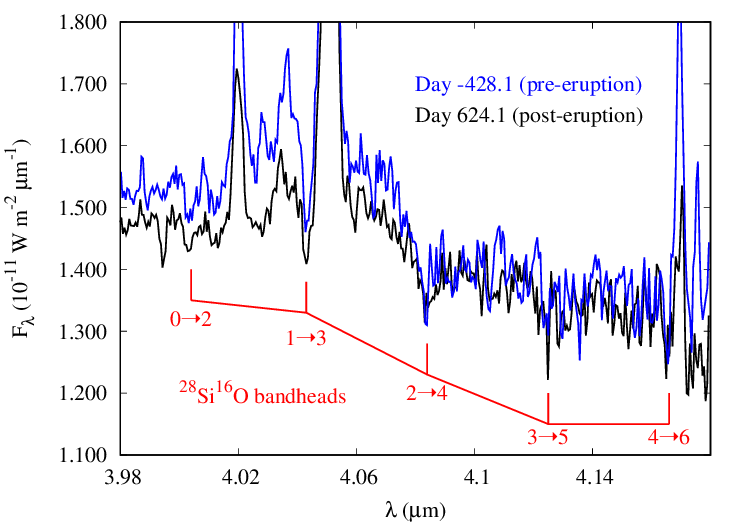}
 \caption{The IRTF spectra in the region of the first overtone
 SiO bands (marked in red).\label{SiO}}
\end{figure}

There is some evidence for the presence of the 
$^{28}$Si$^{16}$O first overtone band heads at
$\sim4$\mic, both on day --428.1 (before the 2021
eruption) and on day~624.1 (post-2021 eruption;
see Fig.~\ref{SiO}); there is no evidence for 
the presence
of SiO in any other spectra. If present, they seem to be
much weaker than they were in T~CrB in 2019 
\citep{pavlenko20}, seven decades after its 1946 eruption.
However the signal-to-noise in the 2021 data is such that
Si isotopic ratios can not be determined. However, high 
resolution IR spectra in the 4\mic\ range should help determine 
isotopic ratios, as was done for T~CrB \citep{pavlenko20}.

\section{Conclusions}
We have presented NIR spectra of \rso, immediately before,
during, and after its 2021 eruption. Our conclusions are as follows:
\begin{enumerate}
\item emission lines observed prior to the eruption most likely
arise in the red giant wind, illuminated by the white dwarf and
the accretion disc;
\item during the eruption, there is evidence for a (relatively)
cool gas at 8900~K, resulting from flash-ionisation of the red
giant wind by hard radiation from the eruption;
\item there are several coronal emission lines in the NIR
during the first month of the eruption; the implied temperature
of the coronal gas is $\sim10^6$~K;
 \item there is no evidence for rapid variability
 in the \ion{He}{i} 1.0833\mic\ line during the SSS phase.
\end{enumerate}

\section*{Acknowledgements}
{We thank the referee for their very careful and
thorough review; their comments have resulted
in considerable improvements to the paper.}
This paper is based on observations obtained under IRTF
program 2021A012, and 2022A04, and the authors thank 
the IRTF staff for their assistance  at the telescope.
CEW acknowledges support from NASA.
SS acknowledges partial support from a NASA Emerging
Worlds grant to ASU (80NSSC22K0361), as well as support
from his ASU Regents’ Professorship.
KLP acknowledges support from the UK Space Agency. 
We acknowledge with thanks the variable star observations
from the \textit{AAVSO International Database} contributed
by observers worldwide and used in this research.

\section*{Data availability}
The IRTF data are available from the IRTF archive\\
http://irtfweb.ifa.hawaii.edu/research/irtf\_data\_archive.php\,.
The Swift data are available from \\
https://www.swift.ac.uk/swift\_live\,.

\appendix

\section{Emission line fluxes}
\begin{landscape}

\begin{table*}
\caption{Dereddened line fluxes. Where the line structure is
comples, fluxes are listed for the strongest, central, component.
Fluxes, in W~m$^{-2}$, are given in the form X.YZ(--EE) = X.YZ$\times10^{-\rm{EE}}$.
Upper limits are $3\sigma$. \label{all-data-tab}}
\begin{tabular}{ccccccccc} 
20200610 & 20210820 & 20210911 & 20220516 & 20220620 & 20220710 & 20230424 & ID & $\lambda$\\
-428.1 & 11.7 & 31.7 & 281.1 & 315.9 & 335.3 & 624.1 & & (\mic)\\ \hline
3.33(--15)&&2.93(--13)&5.06(--15)&3.46(--15)&2.93(--15)&1.64(-15)&\ion{He}{I}&0.7067\\
$<4.5(-16)$&1.68(-14)&1.61(--14)&$<5.1(-16)$&$<3(-15)$&$<1.2(-15)$&&\fion{S}{xii}&0.7613\\
1.31(--15)&4.79(--13)&&&&&&\ion{O}{i}&0.7776\\
&1.54(--13)&6.16(--14)&$<5.7(-16)$&3.80(--16)&$<1.2(-15)$&&\fion{Fe}{xi}&0.7894\\
1.01(--14)&3.91(--12)&1.00(--12)&4.19(--15)&4.59(--15)&3.77(--15)&4.82(--15)&\ion{O}{i}&0.8449\\
4.91(--15)&2.48(--13)&5.84(--14)&3.37(--15)&3.97(--15)&3.41(--15)&4.44(--15)&\ion{Ca}{ii}&0.8500\\
4.33(--15)&1.87(--13)&2.97(--14)&2.15(--15)&3.74(--15)&2.43(--15)&3.48(--15)&\ion{Ca}{ii}&0.8544\\
4.67(--15)&5.63(-13)&1.25(--13)&2.97(--15)&3.79(--15)&3.08(--15)&3.51(--15)&\ion{Ca}{ii}&0.8665\\
&2.74(--13)&6.18(--14)&$<6.3(-16)$&3.90(--16)$^*$&$<1.1(-15)$&&\fion{S}{ix}&0.9391\\
&1.80(--15)&6.08(--14)&1.03(--15)$^*$&$<5.4(-16)$&$<3.6(-17)$&&\fion{S}{viii}&0.9914\\
4.19(--15)&9.69(--13)&1.84(--13)&2.61(--15)&2.21(--15)&1.75(--15)&2.44(--15)&\ion{H}{i} 7--3&1.0052\\
&$<5.4(-15)$&1.84(--13)&3.04(--15)$^*$&&&&\fion{Ar}{xiii}&1.0143\\
&&7.08(--15)&3.92(--16)&3.87(--16)&&&\fion{P}{xi}&1.031\\
&4.33(--13)&2.53(--14)&$<2.5(-16)$&$<1.6(-15)$&$<5.4(-16)$&&\fion{Fe}{xiii}&1.075\\
8.59(--15)&&3.40(--12)&2.52(--14)&1.69(--14)&1.37(--14)&1.09(--14)&\ion{He}{i}&1.0833\\
5.77(--15)&2.71(--12)&5.54(--13)&2.16(--15)&3.09(--15)&2.76(--15)&4.04(--15)&\ion{H}{i} 6--3&1.0941\\
7.10(--15)&2.36(--12)&5.63(--13)&3.94(--15)&3.17(--15)&2.81(--15)&4.19(--15)&\ion{O}{i}&1.129\\
1.70(--16)&9.30(--14)&5.37(--14)&3.72(--16)&1.65(--16)&$<3(-16)$&3.37(--16)&\fion{S}{ix}&1.2535\\
&&&&3.66(--16)&&&\fion{P}{viii}&1.2784\\
&$<1.4(-13)$&$<2.4(-15)$&&&5.46(--16)&&\fion{Cr}{ix}&1.2786\\
9.70(--15)&2.36(--12)&4.75(--13)&4.66(--15)&4.39(--15)&3.60(--15)&5.42(--15)&\ion{H}{i} 5--3&1.2822\\
3.42(--16)&8.17(--14)&5.88(--14)&&&&&\ion{O}{i}&1.3168\\
$<3.3(-16)$&$<1.5(-14)$&2.47(--14)&$<4.6(-16)$&$<5.7(-16)$&&$<7.5(-16)$&\fion{Si}{x}&1.4309\\
$<3.3(-16)$&$<1.4(-14)$&&$<1.1(-15)$&$<4.8(-16)$&$<2.2(-16)$&$<5.7(-16)$&\fion{Si}{ix}&1.5599\\
&&3.08(--14)&4.81(--16)&4.53(--16)&3.64(--16)&$<5.7(-16)$&\fion{P}{viii}&1.7361\\
1.44(--15)&1.45(--13)&1.73(--14)&&&&6.27(--16)&\ion{H}{i} 9--4&1.8179\\
8.39(--15)&3.80(--12)&1.09(--12)&4.39(--15)&4.67(--15)&&5.64(--15)&\ion{H}{i} 4--3&1.8755\\
3.89(--16) &2.84(--13)&4.69(--14)&&&&3.66(--16)&\ion{H}{i} 8--4&1.9461\\
8.75(--16)&$<2.1(-15)$&3.56(--14)&$<4.6(-16)$&$<8.4(-16)$&$<3.6(-17)$&4.28(--16)&\fion{Si}{vi}&1.965\\
&2.46(--14)&2.62(--15)&&$<3.6(-16)$&$<1.4(-16)$&$<1.5(-16)$&\fion{Al}{ix}&2.045\\
7.67(--16)&3.89(--13)&6.33(--14)&7.05(--16)&2.77(--16)&$<1.4(-16)$&7.45(--16)&\ion{He}{i}&2.0587\\
1.51(--15)&2.90(--13)&6.57(--14)&8.04(--16)&6.83(--16)&5.83(--16)&8.86(--16)&\ion{H}{i} 7--4&2.1661\\
&$<8.4(-16)$&3.30(--16)&$<2.6(-16)$&&$<2.0(-16)$&$<2.3(-16)$&\fion{Ca}{viii}&2.321\\
$<3.6(-16)$&9.24(--16)&1.81(--14)&$<2.5(-16)$&$<3.0(-16)$&&$<2.3(-16)$&\fion{Si}{vii}&2.4833\\
\hline\hline
\multicolumn{9}{l}{$^*$The measured wavelength is not a good match to
the line wavelength.} \\
\end{tabular}
\end{table*}

\setcounter{table}{0}

\begin{table*}
\caption{Continued. Dereddened line fluxes. 
Fluxes, in W~m$^{-2}$, are given in the form X.YZ(--EE) = X.YZ$\times10^{-\rm{EE}}$.
Upper limits are $3\sigma$.}
\begin{tabular}{ccccccccc} 
20200610 & 20210820 & 20210911 & 20220516 & 20220620 & 20220710 & 20230424 & ID & $\lambda$\\
-428.1 & 11.7 & 31.7 & 281.1 & 315.9 & 335.3 & 624.1 & & (\mic)\\ \hline
&2.63(--14)&$^\dag$&&&$^\dag$&$<7.7(-17)$&\ion{H}{i} 16--5&2.5261\\
&4.69(--14)&$^\dag$&&&$^\dag$&$<7.6(-17)$&\ion{H}{i} 11--5&2.873\\
&$<3.3(-15)$&$^\dag$&$<1.0(-16)$&$<1.7(-16)$&$^\dag$&$<2.3(-14)$&\fion{Al}{v}&2.9052\\
&$<2.2(-15)$&$^\dag$&&$<1.4(-16)$&$^\dag$&$<1.8(-16)$&\fion{Mg}{viii}&3.028\\
5.83(--16)&7.80(--14)&$^\dag$&&2.16(--16)&$^\dag$&$<1.8(-16)$&\ion{H}{i} 10--5&3.0392\\
&$<1.3(-15)$&$^\dag$&$<7.5(-17)$&$<1.5(-16)$&$^\dag$&$<1.8(-16)$&\fion{Ca}{iv}&3.2067\\
8.14(--16)&9.64(--14)&$^\dag$&&2.32(--16)&$^\dag$&3.69(--16)&\ion{H}{i} 9--5&3.297\\
&2.63(--16)&$^\dag$&$<2.2(-17)$&$<3.0(-17)$&$^\dag$&$<1.6(-17)$&\fion{Al}{vi}&3.6597\\
&1.73(--15)&$^\dag$&&&$^\dag$&&\fion{Al}{viii}&3.69\\
1.30(--16)&&$^\dag$&&&$^\dag$&1.20(--16)&\ion{H}{i} 19--6&3.64593\\
1.18(--16)&2.01(--14)&$^\dag$&&&$^\dag$&9.30(--17)&\ion{H}{i} 18--6&3.6926\\
1.60(--16)&&$^\dag$&&&$^\dag$&&\ion{H}{i} 17--6&3.7494\\
4.34(--16)&1.42(--13)&$^\dag$&&&$^\dag$&2.56(--16)&\ion{H}{i} 8--5&3.7406\\
1.49(--16)&&$^\dag$&&&$^\dag$&&\ion{H}{i} 16--6&3.81945\\
1.52(--16)&&$^\dag$&&&$^\dag$&&\ion{H}{i} 15--6&3.90755\\
2.61(--17)&4.58(--16)&$^\dag$&3.04(--17)&4.67(--17)&$^\dag$&$<1.6(-17)$&\fion{Si}{ix}&3.9357\\
1.62(--15)&5.32(--13)&$^\dag$&7.55(--16)&7.07(--16)&$^\dag$&8.96(--16)&\ion{H}{i} 5--4&4.0523\\
\hline\hline
\multicolumn{9}{l}{$^*$The measured wavelength is not a good match to
the line wavelength.} \\
\multicolumn{9}{l}{$^\dag$Wavelength not covered.}\\
\end{tabular}
\end{table*}
\end{landscape}

\bsp	
\label{lastpage}

\begin{thebibliography}{99}

\bibitem[\protect\citeauthoryear{Abe et al.}{2025}]{abe25} 
Abe K., et al., 2025, A\&A, 695, A152

\bibitem[\protect\citeauthoryear{Anupama \& Miko{\l}ajewska}{1999}]{anupama99}
Anupama G. C., Miko{\l}ajewska J., 1999, A\&A, 344, 177

\bibitem[\protect\citeauthoryear{Anupama}{2008}]{anupama08}
Anupama G. C., 2008, in RS Ophiuchi (2006) and the Recurrent Nova Phenomenon, Astronomical Society of the Pacific Conference Series, eds A. Evans, 
M. F. Bode, T. J. O'Brien, M. J. Darnley, vol. 401, p.~31, San Francisco

\bibitem[\protect\citeauthoryear{Arnaud \& Rothenflug}{1985}]{arnaud85} 
Arnaud M., Rothenflug R., 1985, A\&AS, 60, 425

\bibitem[\protect\citeauthoryear{Asplund, Amarsi \& Grevesse}
{Asplund et al.}{2021}]{asplund21} 
Asplund M., Amarsi A. M., Grevesse N., 2021, A\&A, 653, A141

\bibitem[\protect\citeauthoryear{Azzollini et al.}{2023}]{azzollini23} 
Azzollini A., Shore S. N., Kuin P., Page K. L., 2023,
A\&A, 674, A139

\bibitem[\protect\citeauthoryear{Badnell et al.}{2006}]{badnell06} 
Badnell N. R., et al., 2006, in IAU Symposium 234, 
Planetary Nebulae in our Galaxy and Beyond, ed. M. J. Barlow \& 
R. H. M\'endez (Cambridge: Cambridge University Press), 211

 \bibitem[\protect\citeauthoryear{Baker \& Menzel}{1938}]{baker38}
Baker J. G., Menzel D. H., 1938, ApJ, 88,52
 
\bibitem[\protect\citeauthoryear{Banerjee, Das \& Ashok}{Banerjee et al.}{2009}]{banerjee09} 
Banerjee D. P. K., Das R. K., Ashok N. M., 2009, MNRAS, 399, 357
 
\bibitem[\protect\citeauthoryear{Banerjee et al.}{2023}]{banerjee23}  
Banerjee D. P. K., et al., 2023, ApJ, 954, L16
 
\bibitem[\protect\citeauthoryear{Barry et al.}{2008}]{barry08}
Barry R. K., Mukai K., Sokoloski J. L., Danchi W. C., 
Hachisu I., Evans A., Gehrz R. D., Miko{\l}ajewska J.,
2008, in Evans A., Bode M. F., O'Brien T. J., Darnley M. J., eds,
Astronomical Society of the Pacific Conf. Ser. Vol. 401, RS Ophiuchi (2006) and the Recurrent Nova
Phenomenon. Astron. Soc. Pac., San Francisco, p. 52

\bibitem[\protect\citeauthoryear{Beardmore et al.}{2008}]{beardmore08}
Beardmore A. P., Osborne J. P., Page K. L., Goad M. R., Bode M. F.,
Starrfield S., 2008, in Evans A., Bode M. F., O'Brien T. J., Darnley M. J., eds,
Astronomical Society of the Pacific Conf. Ser. Vol. 401, RS Ophiuchi (2006) and the Recurrent Nova
Phenomenon. Astron. Soc. Pac., San Francisco, p. 296

\bibitem[\protect\citeauthoryear{Beck}{2021}]{beck21}
{Beck S., 2021, AAVSO Alert Notice \#752}

\bibitem[\protect\citeauthoryear{Bell et al.}{2024}]{bell24}
Bell G. S., Berry D. S., Graves S. F., Currie M. J.,
Draper P. W., 2024, in Astromical Data Analysis Software
and Systems XXXI, eds B. V. Hugo, R. Van Rooyen,
O. M. Smirnov, Astronomical Society of the Pacific
Conference Series, 535, 455

\bibitem[\protect\citeauthoryear{Bhatia \& Kastner}{1995}]{bhatia95}
Bhatia A. K., Kastner S. O., 1995, ApJS, 96, 325

\bibitem[\protect\citeauthoryear{Bode \& Kahn}{1985}]{bode85}
Bode M. F., Kahn F. D., 1985, MNRAS, 217, 205

\bibitem[\protect\citeauthoryear{Bode}{1987}]{bode87}
Bode M. F., 1987, editor, RS Ophiuchi (1985) and the recurrent
nova phenomenon, VNU Science Press, Utrecht

\bibitem[\protect\citeauthoryear{Bode et al.,}{2006}]{bode06}
Bode M. F., et al., 2006, IAUC \#8761, \#1

\bibitem[\protect\citeauthoryear{Boffin \& Merc}{2025}]{boffin25}
Boffin H., Merc J., 2025, A\&A, 701, A151

\bibitem[\protect\citeauthoryear{Bollimpalli, Hameury \& Lasota}{2018}]{bollimpalli18} 
Bollimpalli D. A., Hameury J.-M., Lasota J.-P., 2018,
MNRAS, 481, 5422

\bibitem[\protect\citeauthoryear{Bowen}{1947}]{bowen47}
Bowen I. S., 1947, PASP, 59, 196

\bibitem[\protect\citeauthoryear{Brandi et al.}{2009}]{brandi09}
Brandi E., Quiroga C., Milo{\l}ajewska J., Ferrer O. E., 
Garc{\'i}a L. G., 2009, A\&A, 497, 815

\bibitem[\protect\citeauthoryear{Breeveld et al.}{2011}]
{breeveld11}
Breeveld A. A., Landsman W., Holland S. T., Roming P., 
Kuin N. P. M., Page M. J., 2011, in Gamma Ray Bursts, ed J. E.
McEnery, American Institute of Physics Conference Proceedings, 
Vol.~1358, 373

\bibitem[\protect\citeauthoryear{Cardelli, Clayton \& Mathis}{Cardelli et al.}{1989}]{cardelli89} 
Cardelli J. A., Clayton G. C., Mathis J. S., 1989, ApJ, 345, 245

\bibitem[\protect\citeauthoryear{Cheung et al.}{2022}]{cheung22}
Cheung C. C., et al., 2025, ApJ, 935, 44

\bibitem[\protect\citeauthoryear{Chomiuk, Metzger \& Shen}
{Chomiuk et al.}{2021}]{chomiuk21} 
Chomiuk L., Metzger B. D., Shen K. J., 2021, ARA\&A, 59, 391

\bibitem[\protect\citeauthoryear{Cox}{2000}]{cox00}
Cox A. N., 2000, Allen's Astrophysical Quantities, 4th edition. 
AIP Press; Springer, New York

\bibitem[\protect\citeauthoryear{Currie et al.}{2014}]{currie14}
Currie M. J., Berry D. S., jennes T., Gibb A. G., Bell G.
S., Draper P. W., 2014, in
Astronomical Data Analysis Software and Systems XXIII.
eds N. Manset, P. Forshay, Astronomical Society of the 
Pacific Conference Series, vol 485, p.391

\bibitem[\protect\citeauthoryear{Cushing, Vacca \& Rayner}{Cushing et al.}{2004}]{cushing04} 
Cushing M. C., Vacca W. D., Rayner J. T., 2004, PAstronomical Society of the Pacific, 116, 362

\bibitem[\protect\citeauthoryear{Cushing, Rayner \& Vacca}{2005}]{cushing05}
Cushing M. C., Rayner  J. T,  Vacca W. D., 2005, ApJ, 623, 1115


\bibitem[\protect\citeauthoryear{Das, Banerjee \& Ashok}{Das et al.}{2006}]{das06}
Das R., Banerjee D. P. K., Ashok N. M., 2006,
ApJ, 653, L141

\bibitem[\protect\citeauthoryear{de Ruiter et al.}{2023}]{deruiter23} 
de Ruiter I., Nyamai M. M., Rowlinson A., Wijers R. A. M. J.,
O'Brien T. J., Williams D. R. A., Woudt P., 2023, MNRAS, 523, 132

\bibitem[\protect\citeauthoryear{Eggleton}{1983}]{eggleton83} 
Eggleton P. P., 1983, ApJ, 268, 368

\bibitem[\protect\citeauthoryear{Evans et al.}{1988}]{evans88}
Evans A., Callus C. M., Albinson J. S., Whitelock P. A., 
Glass I. S., Carter B., Roberts G., 1988, MNRAS, 234, 755

\bibitem[\protect\citeauthoryear{Evans et al.}{2007a}]{evans07a}
Evans A., et al., 2007a, MNRAS, 374, L1

\bibitem[\protect\citeauthoryear{Evans et al.}{2007b}]{evans07b}
Evans A., et al., 2007b, ApJ, 663, L29

\bibitem[\protect\citeauthoryear{Evans et al.}{2007c}]{evans07c}
Evans A., et al., 2007c, ApJ, 671, L157

\bibitem[\protect\citeauthoryear{Evans et al.}{2008}]{evans08}
Evans A.,  Bode M. F.,  O'Brien T. J.,  Darnley M. J., 2008
editors, RS Ophiuchi (2006) and the Recurrent Nova Phenomenon,
Astronomical Society of the Pacific Conference Series, Vol. 401, 
San Francisco

\bibitem[\protect\citeauthoryear{Evans et al.}{2022}]{evans22} 
Evans A., Geballe T. R., Woodward C. E., Banerjee D. P. K.,
Gehrz R. D., Starrfield S., Shahbandeh M., 2022, MNRAS, 517, 6077

\bibitem[\protect\citeauthoryear{Evans et al.}{2023}]{evans23} 
Evans A., et al., 2023, MNRAS, 522, 4841

\bibitem[\protect\citeauthoryear{Geary \& Amorim}{2021}]{geary21}
{Geary K., Amorim A., 2021, CBET \#5013}

\bibitem[\protect\citeauthoryear{Gehrels et al.}{2004}]{gehrels04}
Gehrels N., et al., 2004, ApJ, 611, 1005

\bibitem[\protect\citeauthoryear{Gehrz et al.}{2008}]{gehrz08} 
Gehrz R. D., et al., 2008, ApJ, 627, 1167

\bibitem[\protect\citeauthoryear{Greenhouse et al.}{1990}]{greenhouse90} 
Greenhouse M. A., Grasdalen G. L., Woodward C. E., Benson J., 
Gehrz R. D., Rosenthal E., Skrutskie M. F., 1990, ApJ, 352, 307


\bibitem[\protect\citeauthoryear{Habtie \& Das}{2025}]{habtie25}
Habtie G. R., Das R., 2025, MNRAS, 537, 2046

\bibitem[\protect\citeauthoryear{Helton et al.}{2012}]{helton12} 
Helton L. A., et al., 2012, ApJ, 755, 37

\bibitem[\protect\citeauthoryear{Herbig \& Soderblom}{1980}]{herbig80}
Herbig G. H., Soderblom D. R., 1980, ApJ, 242, 628

\bibitem[\protect\citeauthoryear{Hummer et al.}{1993}]{hummer93} 
Hummer D. G., et al., 1993, A\&A, 279, 298

\bibitem[\protect\citeauthoryear{Islam, Mukai \& Sokoloski}{Islam et al.}{2024}]{islam24} 
Islam N., Mukai K., Sokoloski J. L., 2024, ApJ, 960, 125

\bibitem[\protect\citeauthoryear{Karzas \& Latter}{1961}]{karzas61} 
Karzas W. J., Latter R., 1961, ApJS, 6, 167

\bibitem[\protect\citeauthoryear{Kastner \& Bhatia}{1995}]{kastner95}
Kastner S. O., Bhatia A. K., 1995, ApJ, 439, 346

\bibitem[\protect\citeauthoryear{Lico et al.}{2024}]{lico24} 
Lico R., et al., 2024, A\&A, 692, A107

\bibitem[\protect\citeauthoryear{Merc \& Boffin}{2025}]{merc25}
Merc J., Boffin H., 2025, A\&A, 695, A61

\bibitem[\protect\citeauthoryear{Merc et al.}{2025}]{merc25+}
{Merc J., et al., 2025, MNRAS, 541, L14}

\bibitem[\protect\citeauthoryear{Miko{\l}ajewska \& Shara}{2017}]{mikolajewska17}
{Miko{\l}ajewska J., Shara M. M., 2017, ApJ, 847, 6}

\bibitem[\protect\citeauthoryear{Munari \& Tabacco}{2022}]{munari22} 
{Munari U., Tabacco F., 2022, RNAAS, 6,103}

\bibitem[\protect\citeauthoryear{Munari \& Valisa}{2021}]{munari21} 
Munari U., Valisa P., 2021, arXiv:2109.01101

\bibitem[\protect\citeauthoryear{Munari \& Valisa}{2022}]{munari22o} 
Munari U., Valisa P., 2022, arXiv:2203.01378

\bibitem[\protect\citeauthoryear{Munari et al.}{2022}]{munari22r} 
Munrai U., Giroletti M., Marcote B., O'Brien T. J., Veres P., Yang J.,
Williams D. R. A., Woudt P., 2022, A\&A, 666, L6

\bibitem[\protect\citeauthoryear{M\"urset \& Schmid}{1999}]{murset99}
M\"urset U., Schmid H. M., 1999, A\&AS, 137, 473

\bibitem[\protect\citeauthoryear{Nayana et al.}{2024}]{nayana24}
Nayana A. J., et al., 2024, MNRAS, 528, 5528

\bibitem[\protect\citeauthoryear{Nelson et al.}{2008}]{nelson08}
Nelson T., Orio M., Cassinelli J. P., Still M., Leibowitz E., 
Mucciarelli P., 2008, ApJ, 673, 1067

\bibitem[\protect\citeauthoryear{Ness et al.}{2023}]{ness23} 
Ness J.-U., et al., 2024, A\&A, 670, A131

\bibitem[\protect\citeauthoryear{Nikolov et al.}{2023}]{nikolov23}
Nikolov Y., Luna G. J. M., Stoyanov K. A., Borisov G.,
Mukai K., Sokoloski J. L., 2023, A\&A, 679, A150

\bibitem[\protect\citeauthoryear{Orio et al.}{2022}]{orio22} 
Orio M., et al., 2022, ApJ, 938, 34

\bibitem[\protect\citeauthoryear{Orio et al.}{2023}]{orio23} 
Orio M., et al., 2023, ApJ, 955, 37

\bibitem[\protect\citeauthoryear{Osborne et al.}{2011}]{osborne11}
Osborne J. P., et al., 2011, ApJ, 727, 124

\bibitem[\protect\citeauthoryear{Osterbrock \& Ferland}{2006}]{osterbrock06} 
Osterbrock D. E., Ferland G. J., 2006, Astrophysics of Gaseous
Nebulae and Active Galactic Nuclei, 2nd edition, Univ. Sci. 
Books, Sausalito, California

\bibitem[\protect\citeauthoryear{Page et al.}{2022}]{page22} 
Page K. L., et al., 2022, MNRAS, 514, 1557

\bibitem[\protect\citeauthoryear{Pavlenko et al.}{2008}]
{pavlenko08}
Pavlenko Ya. V., et al., 2008, A\&A, 485, 541

\bibitem[\protect\citeauthoryear{Pavlenko et al.}{2010}]
{pavlenko10}
Pavlenko Ya. V., Woodward C. E., Rushton M. T.,
Kaminsky B.,  Evans A., 2010, MNRAS, 404, 206

\bibitem[\protect\citeauthoryear{Pavlenko et al.}{2016}]
{pavlenko16}
Pavlenko Ya. V., Kaminsky B., Rushton M. T., Evans A.,
Woodward C. E., Helton L. A., O'Brien T. J., Jones D.,
Elkin V., 2016, MNRAS, 456, 181

\bibitem[\protect\citeauthoryear{Pavlenko et al.}{2020}]
{pavlenko20}
Pavlenko Ya. V.,  Evans A., Banerjee D. P. K., 
Geballe T. R., Munari U., Gehrz R. D., Woodward C. E., 
Starrfield S., 2016, MNRAS, 456, 181

\bibitem[\protect\citeauthoryear{Persson}{1988}]{persson88} 
{Persson S. E., 1988, PASP, 100, 710}

\bibitem[\protect\citeauthoryear{Phan et al.}{2025}]{phan25} 
Phan V. H. M., Cristofari P., Peretti E., Tatischeff V., Ciardi A.,
2025, arXiv:2504.02043

\bibitem[\protect\citeauthoryear{Planquart, Jorissen \& Van Winckel}
{Planquart et al.}{2025}]{planquart25} 
Planquart L., Jorissen A., Van Winckel H., 2025, A\&A, 694, A85
 
\bibitem[\protect\citeauthoryear{Rayner et al.}{2003}]{rayner03} 
Rayner J. T., et al., 2003, PASP, 115, 362

\bibitem[\protect\citeauthoryear{Rayner, Cushing, \& Vacca}{2009}]{rayner09}
Rayner J. T., Cushing M. C.,  Vacca W. D., 2009, ApJS, 185, 289

\bibitem[\protect\citeauthoryear{Rudy et al.}{2003}]{rudy03} 
Rudy R. J., Dimpfl W. L., Lynch D. K., Mazuk S., Venturini C. C.,
Wilson J. C., Puetter R. C., Perry R. B., 2003, ApJ, 596, 1229

\bibitem[\protect\citeauthoryear{Rushton et al.}{2010}]{rushton10}
Rushton M. T., et al., 2010, MNRAS, 401, 99

\bibitem[\protect\citeauthoryear{Rushton et al.}{2022}]{rushton22}
Rushton M. T., Woodward C. E., Gehrz R. D., Evans A., Kaminsky B., 
Pavlenko Ya. V., Eyres S. P. S., 2022, MNRAS, 517, 2526

\bibitem[\protect\citeauthoryear{Schaefer}{2022}]{schaefer22}
Schaefer B. E., 2022, MNRAS, 517, 6150

\bibitem[\protect\citeauthoryear{Schaefer}{2023}]{schaefer23}
Schaefer B. E., 2023, MNRAS, 524, 3146

\bibitem[\protect\citeauthoryear{Shore et al.}{1996}]{shore96}
Shore S. N., Kenyon S. J., Starrfield S., Sonneborn G.,
1996, ApJ, 456, 717


\bibitem[\protect\citeauthoryear{Snijders}{1987}]{snijders87} 
Snijders M. A. J., 1987, Ap\&SpSci, 130, 243

\bibitem[\protect\citeauthoryear{Somero, Hakala \& Wynn}{Somero et al.}{2017}]{somero17}
Somero A., Hakala P. Wynn G. A., 2017, MNRAS, 464, 2784

\bibitem[\protect\citeauthoryear{Starrfield et al.}{2025}]{starrfield25}
Starrfield S., et al., 2025, ApJ, 982, 89

\bibitem[\protect\citeauthoryear{Storey \& Hummer}{1995}]{storey95} 
Storey P. J., Hummer D. G., 1995, MNRAS, 272, 41

\bibitem[\protect\citeauthoryear{Taguchi, Ueta \& Isogai}{Taguchi et al.}{2021}]{taguchi21}
Taguchi K., Ueta T., Isogai K., 2021, ATel\,\#14838

\bibitem[\protect\citeauthoryear{Tomov et al.}{2023}]{tomov23} 
Tomov N. A., et al., 2023, A\&A, 671, A49

\bibitem[\protect\citeauthoryear{van Belle et al.}{1999}]{vanbelle99} 
van Belle G. T., et al., 1999, AJ, 117, 521

\bibitem[\protect\citeauthoryear{van Hoof}{2018}]{vanhoof18} 
van Hoof P. A. M., 2018, Galaxies, 6, 63

\bibitem[\protect\citeauthoryear{Walder, Folini \& Shore}{Walder et al.}{2008}]{walder08} 
Walder R., Folini D., SHore S. N., 2008, A\&A, 484, L9

\bibitem[\protect\citeauthoryear{Wood, M\"uller \& Harper}
{Wood et al.}{2016}]{wood16} 
Wood B. E., M\"uller H.-R., Harper G. M., 2016, ApJ, 829, 74

\bibitem[\protect\citeauthoryear{Woodward et al.}{2008}]{woodward08}
Woodward, C. E., Helton L. A., Evans, A., van Loon J. Th., 2008,
in Evans A., Bode M. F., O'Brien T. J., Darnley M. J., eds,
Astronomical Society of the Pacific Conf. Ser. Vol. 401, RS Ophiuchi (2006) and the Recurrent Nova
Phenomenon. Astron. Soc. Pac., San Francisco, p. 260

\bibitem[\protect\citeauthoryear{Woodward et al.}{2021}]
{woodward21} 
Woodward C. E., Banerjee D. P. K., Geballe T. R., Page K. L., 
Starrfield S.,  Wagner R. M., 2021, ApJ, 922, L10

\bibitem[\protect\citeauthoryear{Worters et al.}{2007}]{worters07}
Worters H. L., Eyres S. P. S., Bromage G. E., Osborne J. P., 
2007, MNRAS, 379, 1557

\bibitem[\protect\citeauthoryear{Wynn}{2008}]{wynn08}
Wynn G., 2008, in Evans A., Bode M. F., O'Brien T. J.,
Darnley M. J., eds, Astronomical Society of the Pacific
Conf. Ser. Vol. 401, RS Ophiuchi (2006) and the
Recurrent Nova Phenomenon. Astron. Soc. Pac., San
Francisco, p. 73
%

\end{thebibliography}
\end{document}